\documentclass[twocolumn, english, prd, aps, showpacs, amsmath, amssymb, floatfix, nofootinbib]{revtex4}

\pdfoutput=1
\usepackage[pdftex]{graphicx}
\usepackage[pdftex]{color}

\def\lsim{\mathrel{\raise.3ex\hbox{$<$\kern-.75em\lower1ex\hbox{$\sim$}}}}
\def\gsim{\mathrel{\raise.3ex\hbox{$>$\kern-.75em\lower1ex\hbox{$\sim$}}}}
\def\ssf{\scriptscriptstyle f}
\def\ssh{\scriptscriptstyle h}
\def\ssfh{\scriptscriptstyle fh}

\begin{document}

\title{Accurate Modeling of Weak Lensing with the \MakeLowercase{s}GL Method}

\author{Kimmo Kainulainen} \email{kimmo.kainulainen@jyu.fi}
\author{Valerio Marra} \email{valerio.marra@jyu.fi}

\affiliation{Department of Physics, 
             University of Jyv\"{a}skyl\"{a}, PL 35 (YFL), 
             FIN-40014 Jyv\"{a}skyl\"{a}, Finland}
\affiliation{Helsinki Institute of Physics, 
             University of Helsinki, PL 64, 
             FIN-00014 Helsinki, Finland}

\begin{abstract}

We revise and extend the stochastic approach to cumulative weak lensing (hereafter the sGL method) first introduced in Ref.~\cite{Kainulainen:2009dw}.
Here we include a realistic halo mass function and density profiles to model the distribution of mass between and within galaxies, galaxy groups and galaxy clusters. We also introduce a modeling of the filamentary large-scale structures and a method to embed halos into these structures. We show that the sGL method naturally reproduces the weak lensing results for the Millennium Simulation. The strength of the sGL method is that a numerical code based on it can compute the lensing probability distribution function for a given inhomogeneous model universe in a few seconds. This makes it a useful tool to study how lensing depends on cosmological parameters and its impact on observations. The method can also be used to simulate the effect of a wide array of systematic biases on the observable PDF. As an example we show how simple selection effects may reduce the variance of observed PDF, which could possibly mask opposite effects from very large scale structures.
We also show how a JDEM-like survey could constrain the lensing PDF relative to a given cosmological model.
The updated \mbox{\tt turboGL} code is available at \mbox{turboGL.org}.

\end{abstract}

\keywords{Gravitational Lenses, Inhomogeneous Universe Models, Observational Cosmology}
\pacs{98.62.Sb, 98.65.Dx, 98.80.Es}


\maketitle

\section{Introduction}
\label{sec:intro}

Inhomogeneities in the large-scale matter distribution can in many ways affect the light signals coming from very distant objects. These effects need to be understood well if we want to map the expansion history and determine the composition of the universe to a high precision from cosmolgical observations. In particular the evidence for dark energy in the current cosmological concordance model is heavily based on the analysis of the apparent magnitudes of distant type Ia supernovae (SNe)~\cite{Kim:2003mq,Kowalski:2008ez}. Inhomogeneities can affect the observed SNe magnitude-redshift relation for example through gravitational lensing, in a way which essentially depends on size and composition of the structures through which light passes on its way from source to observer.

The fundamental quantity describing this statistical magnification is the lensing probability distribution function (PDF). It is not currently possible to extract the lensing PDF from the observational data and we have to resort to theoretical models. Two possible alternatives have been followed in the literature. A first approach (e.g.~Ref.~\cite{Valageas:1999ir, Munshi:1999qw, Wang:2002qc, Das:2005yb}) relates a ``universal'' form of the lensing PDF to the variance of the convergence, which in turn is fixed by the amplitude of the power spectrum, $\sigma_{8}$. Moreover the coefficients of the proposed PDF are trained on some specific N-body simulations. A second approach (e.g.~Ref.~\cite{Holz:1997ic, Holz:2004xx}) is to build {\em ab-initio} a model for the inhomogeneous universe and directly compute the relative lensing PDF, usually through time-consuming ray-tracing techniques. The flexibility of this method is therefore penalized by the increased computational time.

In Ref.~\cite{Kainulainen:2009dw} we introduced a stochastic approach to cumulative weak lensing (hereafter sGL method) which combines the flexibility in modeling with a fast performance in obtaining the lensing PDF. The speed gain is actually a sine-qua-non for likelihood approaches, in which one needs to scan many thousands different models (see Ref.~\cite{Amendola:2010ub}). The sGL method is based on the weak lensing approximation and generating stochastic configurations of inhomogeneities along the line of sight. The major improvements introduced here are the use of a realistic halo mass function to determine the halo mass spectrum and the incorporation of large-scale structures in the form of filaments. The improved modeling together with the flexibility to include a wide array of systematic biases and selection effects makes the sGL method a powerful and comprehensive tool to study the impact of lensing on observations. 

We show in particular that the sGL method, endowed with the new array of inhomogeneities, naturally and accurately reproduces the lensing PDF of the Millenium Simulation~\cite{Hilbert:2007ny, Hilbert:2007jd}. We also study a simple selection effect model and show that selection biases can reduce the variance of the observable PDF. Such reduction could at least partly cancel the opposite effect coming from large scale inhomogeneities, masking their effect on the observable PDF.
We also show how a JDEM-like survey could constrain the lensing PDF relative to a given cosmological model.
Along with this paper, we release an updated version of the \mbox{\tt turboGL} package, which is a simple and very fast Mathematica implementation of the sGL method \cite{turboGL}.

This paper is organized as follows. In Section~\ref{sup} we introduce the cosmological background, the generic layout of inhomogeneities and review the basic formalism needed to compute the weak lensing convergence. In Section~\ref{sec:inhoprop} we derive the halo mass function and the halo density profiles and define the precise modeling of filaments. In Section~\ref{hmf} we present the revised and extended sGL method. The exact discretization of the model parameters, which is a crucial step in the sGL model building, is explained in Section~\ref{mbin} and in Section~\ref{confi} we explain how the realistic structures where halos are confined in filaments are modelled in the sGL method. Section~\ref{results} contains our numerical results including the comparison with the cosmology of the Millennium Simulation~\cite{Springel:2005nw} and, finally, in Section~\ref{conco} we will give our conclusions.

\section{Setup \label{sup}}

\subsection{Cosmological Background \label{bkg}} 

We consider homogeneous and isotropic Friedmann-Lema\^{i}tre-Robertson-Walker (FLRW) background solutions to Einstein's equations, whose metric can be written as:
\begin{equation} \label{metric}
ds^{2}=-c^{2}dt^{2}+a^{2}(t) \left[ dr^{2}+f_{K}^{2}(r) d\Omega^{2} \right] \, ,
\end{equation}
where $d\Omega^{2}=d \theta ^{2}+\sin^{2} \theta d\phi^{2}$ and
\begin{equation}
f_K(r) = \left\{
  \begin{array}{ll}
    K^{-1/2}\sin(K^{1/2}r) & \; K>0 \\
    r & \; K=0 \\
    (-K)^{-1/2}\sinh[(-K)^{1/2}r] & \; K<0 \\
  \end{array}\right.\; ,
\end{equation}
where $K/a^{2}(t)$ is the spatial curvature of any $t-$slice.

In particular we will focus on $w$CDM models whose Hubble expansion rate depends on redshift according to:
\begin{eqnarray}
{H^2(z) \over H_{0}^{2} }&\equiv& E^{2}(z)= \Omega_{Q0} \, (1+z)^{3 q(z)} \label{hhh} \\
 &+& \Omega_{K0} \, (1+z)^{2}  + \Omega_{M0}  \, (1+z)^{3} +\Omega_{R0} \, (1+z)^{4} \, ,  \nonumber
\end{eqnarray}
where $q(z)$ is given by:
\begin{equation}
q(z)= {1 \over \ln (1+z)} \int_{0}^{z}{1+w(z') \over 1+z'}dz' \, .
\end{equation}
Here $w(z)$ could be taken to follow {\em e.g.}~the parameterization~\cite{depar}:
\begin{equation}
w(z)=w_{0}+ w_a {z \over 1+z} \, ,
\end{equation}
for which:
\begin{equation}
q(z)= 1+w_{0}+w_{a}- {w_{a} \, z \over (1+z)\ln (1+z)} \, .
\end{equation}
For a constant equation of state $w(z)=w_{0}$, the latter reduces to $q(z)=1+w_{0}$.

$\Omega_{Q0}$, $\Omega_{M0}$ and $\Omega_{R0}$ are the present-day density parameters of dark energy, matter and radiation and $\Omega_{K0}=1-\Omega_{Q0} - \Omega_{M0} - \Omega_{R0}$ represents the spatial-curvature contribution to the Friedmann equation.
Introducing the Hubble radius $L_{H}=c/H$ and the spatial-curvature radius $L_{K}=a/(\pm K)^{1/2}$, we have at any time the relation $\Omega_{K}=-L_{H}^{2}/L_{K}^{2}$.
We will also need the matter and dark energy density parameters at a given time or redshift:
\begin{eqnarray}
\Omega_{M}(z) &=& \Omega_{M0} {(1+z)^{3} \over E^{2}(z)} \, , \\
\Omega_{Q}(z) &=& \Omega_{Q0} {(1+z)^{3 q(z)}   \over E^{2}(z)} \, .
\end{eqnarray}
Throughout this paper, the subscript $0$ will denote the present-day values of the quantities.
For example the critical density today is $\rho_{C0}=3 H_{0}^{2}/8 \pi G$ where $H_{0}=100  h$ km s$^{-1}$ Mpc$^{-1}$, while $\rho_{C}=3 H^{2}/8 \pi G$ is the critical density at any time.

Substituting in Eq.~(\ref{hhh}) $H=\dot{a}(t)/a(t)$ and $1+z=a_{0}/a(t)$ we obtain the equation we have to solve in order to find the time evolution of the scale factor $a(t)$, the only dynamical variable in an FLRW model.
We fix the radiation density parameter to $\Omega_{R0} =4.2 \cdot 10^{-5} h^{-2}$. Moreover $\Omega_{M}=\Omega_{DM}+\Omega_{B}$, that is, dark and baryonic matter contribute together to the total matter density. The line-of-sight and transverse comoving distances are:
\begin{eqnarray}
d_{C \parallel}(z)& \equiv & r(z)={L_{H0} \over a_{0}} \int_{0}^{z}{dz' \over E(z')} \, , \\
d_{C \perp}(z)&=&f_{K}(r(z)) \, ,
\end{eqnarray}
from which we find the angular and luminosity distances together with the distance modulus:
\begin{eqnarray}
d_{A}(z) &=& a_{0} (1+z)^{-1} f_{K}(r(z)) \\
d_{L}(z) &=& a_{0} (1+z) f_{K}(r(z)) \\
m(z)&=& 5 \log_{10}{d_L(z) \over 10\textrm{pc}} \,.
\end{eqnarray}
For scales smaller than the spatial-curvature radius $(\pm K)^{1/2} \Delta r = a \Delta r / L_{K} \ll 1$, $f_{K}(\Delta r) \simeq \Delta r$ and we can use the Euclidean geometry.

\subsection{Matter Inhomogeneities \label{mati}} 

The aim of this paper is to compute statistical weak lensing corrections to the measured light intensities, generated by the inhomogeneous matter distribution along the line of sight to a distant object. The basic quantity we need to model is the matter density contrast $\delta_{M}$:
\begin{equation}
\delta_{M}(r,t)= {\rho_{m}(r,t) \over \rho_{M}(t)} -1 \, ,
\label{contra1}
\end{equation}
where the lowercase $\rho_{m}$ indicates the local and inhomogeneous matter field while $\rho_{M} = \Omega_M \, \rho_C$ is the time dependent average mass density. The density contrast directly enters the expression for the weak lensing convergence~\cite{Bartelmann:1999yn}:
\begin{equation}
\kappa  = \int_{0}^{r_{s}} dr \, G(r,r_{s}) \, \rho_{MC} \, \delta_{M} \,, \label{eq:kappa1} 
\end{equation}
where $r_{s}$ is the co-moving position of the source and the integral is along an unperturbed light geodesic. The density $\rho_{MC} \equiv a_0^3 \, \rho_{M0}$ is the constant matter density in a co-moving volume and we defined the auxiliary function\footnote{Note that this definition slightly differs from the one of Ref.~\cite{Kainulainen:2009dw}.}
\begin{equation}
G(r,r_{s})=  \frac{4\pi G}{c^2} \, 
  \frac{f_{K}(r)f_{K}(r_{s}-r)}{f_{K}(r_{s})} \, {1 \over a} \,,
\end{equation}
which gives the optical weight of an inhomogeneity at the comoving radius $r$.
The convergence is related to the shift in the distance modulus by:
\begin{equation} 
\Delta m = 5 \log_{10}\mu^{-1 / 2} \simeq  5 \log_{10}(1-\kappa) \,,
\label{eq:mukappa}
\end{equation}
where $\mu$ is the net magnification and the second-order contribution of the shear has been neglected~\cite{Bartelmann:1999yn,Kainulainen:2009dw}. It is obvious that an accurate statistical modeling of the magnification PDF calls for a detailed description of the inhomogeneous mass distribution.

In this paper we will significantly improve the modeling of the inhomogeneities from our previous work~\cite{Kainulainen:2009dw}, where only single-mass spherical overdensities were considered.
First of all, we improve the modeling of these ``halos" by using a realistic halo mass function $f(M,z)$, which gives the fraction of the total mass in halos of mass $M$ at the redshift $z$. The function $f(M,z)$ is related to the (comoving) number density $n(M,z)$ by:
\begin{equation}
{\rm d}n(M,z) \equiv n(M,z) {\rm d}M = 
          {\rho_{MC} \over M} \, f(M,z) {\rm d}M \, ,
\label{eq:halomassf}
\end{equation}
where we defined ${\rm d}n$ as the number density of halos in the mass range ${\rm d}M$. The halo function is by definition normalized to unity
\begin{equation}
\int f(M,z) {\rm d}M = 1 \,.
\label{eq:fMnorm}
\end{equation}
The idea is of course that halos describe large virialized mass concentrations such as large galaxies, galaxy clusters and superclusters.
Not all matter is confined into virialized halos however. Moreover, only very large mass concentrations play a significant role in our weak lensing analysis: for example the stellar mass in galaxies affects the lensing PDF only at very large magnifications \cite{Hilbert:2007ny, Hilbert:2007jd} where the PDF is close to zero.
Therefore the fraction of mass $\Delta f_H$ concentrated in large virialized halos can be defined by introducing a lower limit to the integral (\ref{eq:fMnorm}):
\begin{equation}
\Delta f_H (z) \equiv \int_{M_{\rm cut}}  f(M,z) {\rm d}M < 1 \,.
\label{eq:def_of_fH}
\end{equation}
Only mass concentrations with $M>M_{\rm cut}$ are treated as halos. The remaining mass is divided into a family of large mass, but low density contrast objects with a fraction $\Delta f_L$ and a uniform component with a fraction $\Delta f_U= \rho_{MU}/\rho_{M}$, such that
\begin{equation}
 \Delta f_H + \Delta f_L +\Delta f_U =  1 \,.
\label{eq:sumrule}
\end{equation}
The low contrast objects are introduced to account for the filamentary structures observed in the large scale structures of the universe. In our analysis they will be modeled by elongated objects with random positions and orientations. The mass in these objects can consist of a smooth unvirialized (dark) matter field and/or of a fine ``dust'' of small virialized objects with $M<M_{\rm cut}$. In weak lensing this distinction does not matter because small halos act effectively as a mean field, with a sizeable contribution only at very large magnifications (see comment above about stellar mass).

For later use we define $\Delta f_{LU} \equiv \Delta f_L +\Delta f_U=1- \Delta f_H$ which gives the total mass fraction not in large virialized halos.
If we only consider virialized masses larger than companion galaxies ($M_{\rm cut} \sim 10^{10} \, h^{-1} M_{\odot}$), then typical values for the concordance model are $\Delta f_{LU} \sim 0.5$ at $z=0$\cite{Springel:2005nw} and $\Delta f_{LU} \sim 0.7$ at $z=1.5$, with a weak dependence on the particular $f(M,z)$ used. Although $\Delta f_{LU}$ is sensitive on $M_{\rm cut}$, the lensing PDF depends only weakly on the cut mass. Moreover, halo functions obtained from N-body simulations are valid above a mass value imposed by the numerical resolution of the simulation itself and so the use of $M_{\rm cut}$ is also necessary in this case \cite{Jenkins:2000bv}.

To obtain an as accurate modeling as possible, we will use realistic mass functions and spatial density profiles for the halo distribution. There is less theoretical and observational input to constrain the mass distribution of the filamentary structures or their internal density profiles. We will therefore parametrize the filaments with reasonable assumptions for their lengths and widths and by employing cylindrical nonuniform density profiles.
\begin{figure}
\begin{center}
\includegraphics[width=8.0 cm]{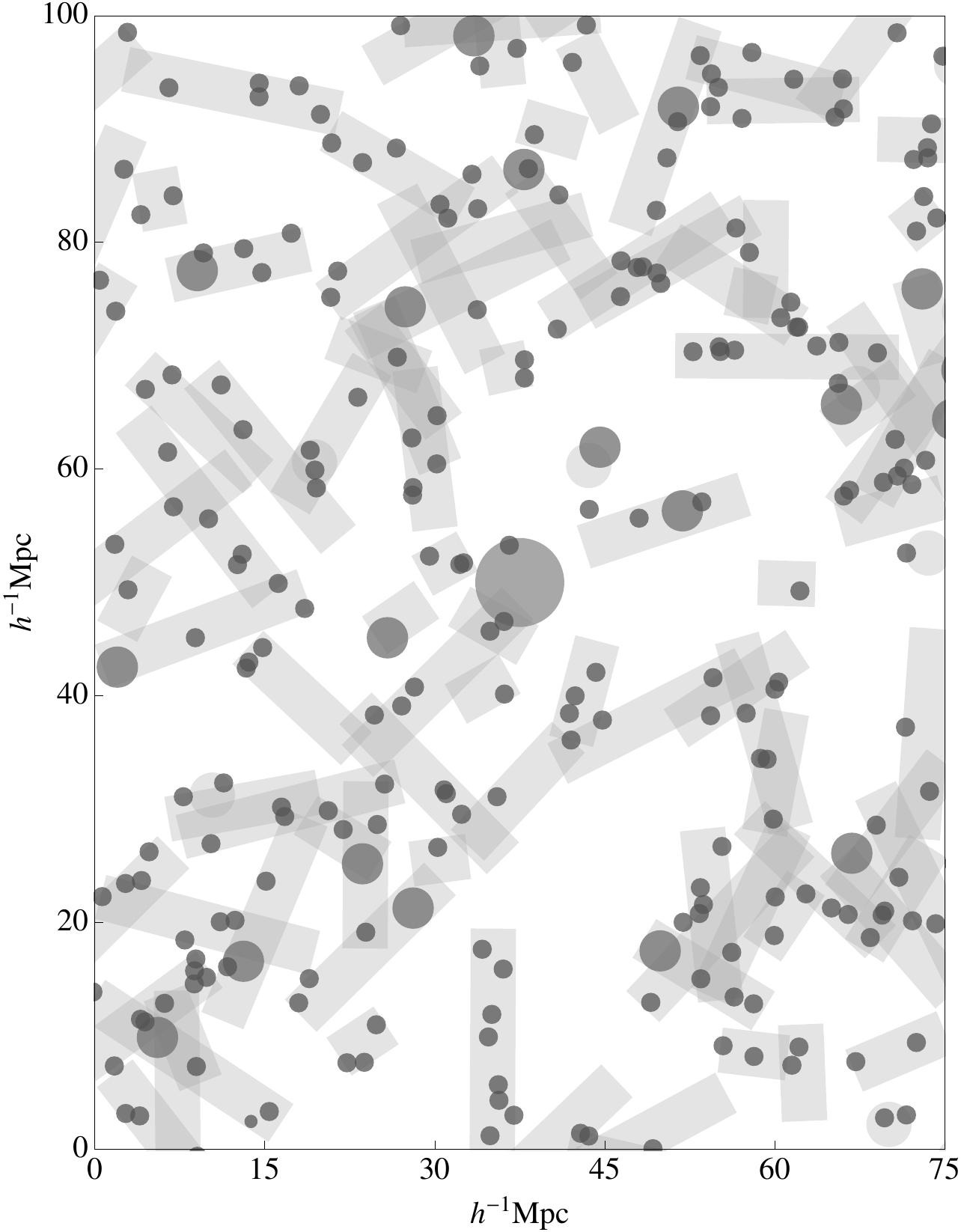}
\caption{Shown is an illustrative-only projection of the random matter density within a $100 h^{-1}$ Mpc thick slice generated by our stochastic model. The shaded disks represent clusters and the shaded cylinders the filamentary dark matter structures. Only large clusters are displayed.}
\label{slice}
\end{center}
\end{figure}
Our modeling allows treating the two families of inhomogeneities independently. Both can be given random spatial distributions, or alternatively all halos can be confined to have random positions in the interiors of randomly distributed cylinders. The latter configuration more closely resembles the observed large scale structures, and an illustration created by a numerical simulation using the {\tt turboGL} package is shown in Fig.~\ref{slice}. 
In Section \ref{discu} we will discuss the power spectrum and the large-scale correlations of our model universe.

We can now formally rewrite Eq.~(\ref{contra1}) for our model universe in which the density distribution is given by~$\rho_{m}=\sum_{j} \rho_{j}+\rho_{MU}$:
\begin{equation} 
\delta_{M} = {\sum_{j} M_{j} \varphi_{j} \over \rho_{MC}} +\Delta f_{U} -1 \, ,
\label{contra2}
\end{equation}
where the index $j$ labels all the inhomogeneities, we used Eq.~(\ref{eq:sumrule}) and we defined $\rho_{j} \equiv a^{-3} \, M_{j} \, \varphi_{j}$, so that both virialized halos and the unvirialized objects are described by a generic reduced density profile $\varphi_j$ which satisfies the normalization $\int_{V} \varphi \, dV=1$.\footnote{Note that this definition slightly differs from the one of Ref.~\cite{Kainulainen:2009dw}.}
The lensing convergence we are interested in can now be written as
\begin{equation}
\kappa \equiv \kappa_{HL} + \kappa_U + \kappa_E  \,,
\label{eq:kappatot}
\end{equation}
where the positive contribution due to the inhomogeneities (halos and low contrast objects) is
\begin{equation}
\kappa_{HL} = \int_{0}^{r_{s}} dr  \, G(r,r_{s})  \sum_{j} M_{j} \varphi_{j} \,,
\label{eq:kappaIH}
\end{equation}
the also positive contribution due to the uniformly distributed matter is
\begin{equation}
\kappa_{U}=\rho_{MC}  \int_{0}^{r_{s}} dr \, G(r,r_{s}) \, \Delta f_{U}(z(r)) \,,
 \label{keun}
 \end{equation}
and the negative empty beam convergence is:
\begin{equation}
\kappa_{E}=-\rho_{MC}  \int_{0}^{r_{s}} dr \, G(r,r_{s}) \,.
 \label{kem}
 \end{equation}
A light ray that misses all the inhomogeneities $\varphi_{j}$ will experience a negative total convergence $\kappa_{UE} = \kappa_U + \kappa_E$, which gives the maximum possible demagnification in a given model universe.
In an exactly homogeneous FLRW model the two contributions $\kappa_{HL}$ and $\kappa_ {UE}$ cancel and there is no net lensing. In an inhomogeneous universe, on the other hand, a light ray encounters positive or negative density contrasts, and its intensity will be magnified or demagnified, respectively. The essence of the sGL method is finding a simple statistical expression for the probability distribution of the quantity $\kappa_{HL}$ in the inhomogeneous universe described above.
 
For a discussion about the validity of the weak lensing approximation within our setup see Ref.~\cite{Kainulainen:2009dw}, where it was shown that the error introduced is $\lesssim 5$\%. We stress again that the lensing caused by stellar mass in galaxies is negligible in the weak lensing regime \cite{Hilbert:2007ny, Hilbert:2007jd} and so we can focus on just modeling the dark matter distribution in the universe.
Also, we will treat the inhomogeneities as perturbations over the background metric of Eq.~(\ref{metric}). In particular we will assume that redshifts can be related to comoving distances through the latter metric. See Ref.~\cite{Kantowski:1969, SC-re, Biswas:2007gi, Brouzakis:2008uw} for a discussion of redshift effects.

In the next section we will give the accurate modeling of the inhomogeneities.  We begin by introducing the halo mass function and the detailed halo profiles, and then move on to describe the precise modeling of the cylindrical filaments.

\section{Halo and filament properties}
\label{sec:inhoprop}

We begin by explaining our dark matter halo modeling. The two main concepts here are the halo mass function $f(M,z)$ giving the normalized distribution of the halos as a function of their mass and redshift, and the dark matter density profile within each individual halo. Both quantities are essential for an accurate modeling of the weak lensing effects by inhomogeneities. We shall begin with the halo mass function introduced above in Eq.~(\ref{eq:halomassf}).

\subsection{Halo Mass Function} 

The halo mass function acquires an approximate universality when expressed with respect to the variance of the mass fluctuations on a comoving scale $r$ at a given time or redshift, $\Delta(r,z)$. Relating the comoving scale $r$ to the mass scale by $M = {4 \pi \over 3}  r^{3}  \, \rho_{MC}$, we can define the variance in a given mass scale by $\Delta^{2}(M,z) \equiv \Delta^{2}(r(M),z)$.
The variance $\Delta(r,z)$ can be computed from the power spectrum:
\begin{equation} \label{d2r}
\Delta^{2}(r,z) \equiv \left( {\delta M \over M} \right)^{2}=\int_{0}^{\infty} {dk \over k} \Delta^{2}(k,z) W^{2}(k  r) \, ,
\end{equation}
where $W(k  r)$ is the Fourier transform of the chosen (top-hat in this work) window function and the dimensionless power spectrum extrapolated using linear theory to the redshift $z$ is:
\begin{eqnarray} \label{d2k}
\Delta^{2}(k,z) &\equiv& {k^{3} \over 2 \pi^{2}} P(k, z) \\
&=& \delta_{H0}^{2} \left( {c k \over a_{0}H_{0}} \right )^{3+n_{s}} T^{2}(k/a_{0}) D^{2}(z) \, . \nonumber
\end{eqnarray}
Here $n_{s}$ is the spectral index and $\delta_{H0}$ is the amplitude of perturbations on the horizon scale today, which we fix by requiring:
\begin{equation}
\Delta(r=8/a_{0}h \textrm{ Mpc},z=0)= \sigma_{8} \, ,
\end{equation}
where the value of $\sigma_{8}$ is estimated by cluster abundance constraints \cite{Rozo:2009jj}. $D(z)$ is the linear growth function which describes the growth speed of the linear perturbations in the universe. Fit functions for $D$ could be found, {\em e.g.}, in Ref.~\cite{fitD}, but it can also be easily solved numerically from the equation:
\begin{eqnarray}
D''(z)&+&{1 \over 2} {D'(z) \over 1+z} \left[ \Omega_{M}(z) + (3 w(z) +1) \Omega_{Q}(z) \right] \nonumber \\
&-&{3 \over 2} { D(z) \over (1+z)^{2}} \, \Omega_{M}(z)=0 \,,
\end{eqnarray}
where we have neglected the radiation. As usual, we normalize $D$ to unity at the present time, $D(0)=1$. Finally, for the transfer function $T(k)$ we use the fit provided by the Equations (28-31) of Ref.~\cite{Eisenstein:1997ik}, which accurately reproduces the baryon-induced suppression on the intermediate scales but ignores the acoustic oscillations, which are not relevant for us here.

With $\Delta(M,z)$ given, we can now define our halo mass function. Several different mass functions have been introduced in the literature, but here we will consider the mass function given in Eq.~(B3) of Ref.~\cite{Jenkins:2000bv}:
\begin{equation} \label{jenk}
f_{J}(M,z)= 0.301\,  \exp (-|\ln \Delta(M,z)^{-1}+ 0.64  |^{3.82}) \, ,
\end{equation}
which is valid in the range $-0.5 \le \ln \Delta^{-1} \le 1.0$.
Because of the change of variable, $f_{J}$ is related to our original  definition of $f$ by:
\begin{equation}
f(M,z)= f_{J}(M,z) {d\ln \Delta(M,z)^{-1} \over dM} \, .
\end{equation}

The mass function of Eq.~(\ref{jenk}) is defined relative to a spherical-overdensity (SO) halo finder, and the overdensity used to identify a halo of mass $M$ at redshift $z$ is $\Delta_{SO} =180$ with respect to the mean matter density $\rho_{M}(z)$. The SO finder allows therefore a direct relation between halo mass $M$ and the radius $R_p$:
\begin{equation} \label{mdel}
M = {4 \pi \over 3}  R_{p}^{3}   \, \rho_{M}(z) \, \Delta_{SO} \, .
\end{equation}
The subscript $p$ will denote the proper values of otherwise comoving quantities throughout this paper.  
For example, the comoving halo radius $R$ is related to the proper value by $R_{p}=a \, R$.

A direct relation between the mass and the radius of a cluster is necessary for our sGL method. This is why we prefer mass functions based on SO finders over mass functions based on friends-of-friends halo finders; for the latter an appropriate $\Delta_{SO}$ to be used in Eq.~(\ref{mdel}) is not directly available. Moreover, as shown in Ref.~\cite{Jenkins:2000bv,White:2002at}, the SO(180) halo finder gives a good degree of universality to the mass function. Another mass function which manifests approximate universality with the SO(180) halo finder \cite{White:2002at} is provided by Sheth \& Tormen in Ref.~\cite{Sheth:1999mn}.

\subsection{Halo Profile}
\label{subsec:nfw} 

With the halo mass function given the only missing ingredient is the halo profile which, as said after Eq.~(\ref{contra2}), we describe by means of the reduced halo profile $\varphi_h$. We stress that our halo of a given mass $M$ and redshift $z$ is an avererage representative of the total ensemble of halos, which in reality have some scatter in the density profiles.

We will focus on the Navarro-Frenk-White (NFW) profile \cite{Navarro:1995iw}:
\begin{equation}
{\rho_h \over \rho_{M}} = {\delta_{c} \over (r_{p} / R_{s}) (1+r_{p} / R_{s})^{2} } \, .
\label{eq:NFWprofile}
\end{equation}
Here the scale radius is related to the halo radius by $R_{p}=c \, R_{s}$, where $c$ is the concentration parameter to be defined below. By integrating equation (\ref{eq:NFWprofile}) one finds the total mass $M=4 \pi \, \delta_{c} \, \rho_{M} (R_{p}/c)^{3} [ \ln (1+c) - c/(1+c)]$ which, when combined with Eq.~(\ref{mdel}), gives:
\begin{equation}
\delta_{c} = {\Delta_{SO} \over 3} {c^{3} \over  \ln (1+c) - c/(1+c)} \, .
\end{equation}
This relation fixes $\delta_{c}$ once the concentration parameter is given.
The reduced halo profile $\varphi_h$ in comoving coordinates is then:
\begin{equation} \label{nfw}
\varphi_{h}(r,t) = \left[ 4 \pi  r \left ( {R \over c} +r \right)^{2} \left(  \ln (1+c) - {c \over 1+c} \right) \right]^{-1}  ,
\end{equation}
where the $t$-dependence (as well as $M$-dependence) arises through the redshift dependence of $R$ (Eq.~(\ref{mdel})) and $c$ (Eq.~(\ref{duffy})).

The last ingredient needed to fully specify the NFW profile is a relation between the concentration parameter and the halo mass at a given redshift. For the $\Lambda$CDM model we will use the following fit obtained from numerical simulations~\cite{Duffy:2008pz} satisfying WMAP5 cosmology~\cite{Komatsu:2008hk}:
\begin{equation} \label{duffy}
c(M,z) =  10.14 \, \left( {M \over  M_{R} }\right)^{-0.081} \, (1+z)^{-1.01} \, ,
\end{equation}
where the pivot mass scale is $M_{R}=2  \cdot 10^{12} h^{-1} M_{\odot}$.

Apart from the NFW profile, alternative simple density profiles of interest include:
\begin{eqnarray}
\varphi_{h} &=& {3 \over 4 \pi R^{3}} \phantom{ggggggggggggggggggji} \textrm{Uniform,} \label{uni} \\
\varphi_{h} &=& 0.97^{-1} \, {\exp [-r^{2}/2(R / 3)^{2}] \over (2 \pi)^{3/2} (R/3)^{3}} \phantom{ggg} \textrm{Gaussian,} \label{gau} \\
\varphi_{h} &=& {1 \over 4 \pi r^{2} R} \phantom{ggggggggggggggggggi} \textrm{SIS,} \label{sis}
\end{eqnarray}
where $R$ is again given by Eq.~(\ref{mdel}).
The normalization factor in (\ref{gau}) is necessary to have the correct volume integral of unity and is due to the restriction of the profile between 0 and $R$.
In these cases there are no further parameters to be fixed.

\subsection{Filaments}

We will model dark matter filaments by cylindrical, non-uniform low density objects. Let us stress that throughout this chapter by a ``filament" we refer only to the smoothly distributed dark matter component of a real filament; we shall later explain how virialized halos can be embedded within these objects to form a more realistic model of the true filamentary structures in the full sGL framework.
Keeping the low contrast component and the halos as separate entities in our evaluation is convenient for the modeling and explicitly avoids double counting.

Using Eqs.~(\ref{eq:halomassf}) and (\ref{eq:sumrule}) we can obtain the following equation which relates the comoving filament density $\Delta n_{f}$ and the filament mass $M_{f}$ to the fraction $\Delta f_{LU}$ of mass not confined in virialized halos:
\begin{equation} 
M_{f} \, \Delta n_{f} \equiv \rho_{MC} \,  \beta_{f} \, \Delta f_{LU} \,,
\label{eq:fila}
\end{equation}
where the quantity $\beta_f$ defines the fraction of the total unvirialized mass which is confined in filaments: $\Delta f_{L} = \beta_f \, \Delta f_{LU}$.
Because of the ongoing halo formation $\Delta f_{LU}$ decreases with time and we shall simply assume that $\beta_{f}$ is a constant.

We will assume that all filaments have the same mass, but this assumption can easily be relaxed later by introducing a filament mass function \cite{Shen:2005wd}. We are also neglecting merging of smaller filaments into larger ones, whereby the comoving number density of filaments $\Delta n_{f}$ remains a constant. This assumption seems to be supported by numerical simulations \cite{Springel:2005nw} at least at low redshifts. From Eq.~(\ref{eq:fila}) it then follows that the filament mass $M_{f}$ is not constant, but it gradually decreases as more and more halos virialize.
(The total mass of the full filamentary structure which includes the halos remains on average constant.)

Now we need to specify the precise geometry, dimensions and time-dependent density profile $\varphi_{f}$ for our filaments.  First, we will relate the filament mass to its length and radius at present day according to:
\begin{equation}
M_{f0} = \pi R_{p0}^{2}  L_{p0} \; \rho_{M0} \, \Delta_{f0} \equiv V_{f p0} \, \rho_{M0} \, \Delta_{f0}   \,,
\end{equation}
where $R_{p0}$ and $L_{p0}$ are the present-day filament proper radius and length and $\Delta_{f0}$ is the average overdensity of the filament with respect to the FLRW matter density. Consistent with our assumptions of neglecting filament mergers and using a constant comoving filament density, we will assume that filaments have a constant comoving length. This is again in accordance with the generic picture of a web of filaments of constant comoving length condensing with time seen in numerical simulations. Finally, for simplicity, we will choose our filaments to have a constant proper radius. Summarizing, we will use:
\begin{eqnarray}
L&=& L_{p0}/ a_{0} \,, \nonumber \\
R(t) &=& R_{p0}/a(t) \,.
\end{eqnarray}
The simplest assumption would be to take filaments to have uniform density. In this case the reduced density profile is simply $\bar \varphi_{f}= V_{f}^{-1}$.  However, we will also consider a profile which is uniform along the length of the filament $L$, but which has a gaussian profile in the radial direction. In this case
\begin{equation} \label{filagau}
\varphi_{f}(r,t) = 0.989^{-1} \, {\exp [-r^{2}/2(R / 3)^{2}] \over L \, 2 \pi (R/3)^{2}}  \,,
\end{equation}
where $0 \le r \le R$ and $0 \le l \le L$, and $\varphi_{f}=0$ otherwise.

Finally, let us compute the mass of the ``dressed'' filaments which comprises the ``bare'' filaments until now discussed together with the halos (to be embedded in Section \ref{confi}).
It is easy to find that:
\begin{equation} \label{dressed}
M_{f0}^{D} = V_{f p0} \, \rho_{M0} \,  \left [ \Delta_{f0} \left ( 1 + {\Delta f_{H} \over \Delta f_{L}} \right ) + \Delta f_{U} \right ]  \,,
\end{equation}
where also the contribution from the uniformly distributed matter is included.
Eq.~(\ref{dressed}) is valid if all halos are embedded within the filaments and has to be slightly modified if different levels of confinement are considered.

This completes our description of the {\em objects} in our model. Next we shall have to understand how to simulate their effect on the weak lensing properties of the complete inhomogeneous model universe.

\section{The \MakeLowercase{s}GL Method 
\label{hmf}}

In this section we will review and extend the stochastic gravitational lensing (sGL) method of Ref.~\cite{Kainulainen:2009dw} for computing the probability distribution function of the lens convergence $\kappa$ in the presence of inhomogeneities. The basic quantity to evaluate is the inhomogeneity-induced part $\kappa_{HL} $ in Eq.~(\ref{eq:kappaIH}):
\begin{equation}
\kappa_{HL} (z_s) = \int_0^{r_s}dr \, G(r,r_s) \sum_{j} M_{j} \, \varphi_{j}(|r-r_j|, t_{j})  \,,
\label{eq:kappa3b}
\end{equation}
where $r_s = r(z_s)$. We wish to obtain a probabilistic prediction for this quantity along a line of sight to a source located at $r_s$, through a random distribution of halos and filaments. The problem could be solved by constructing a large comoving volume of statistically distributed objects such as the one shown in Fig.~\ref{slice}, and then computing the distribution of $\kappa_{HL}$ along random directions in this space. However, an equivalent and computationally much more efficient approach is to construct random realizations of object locations along a fixed geodesic and compute the lensing PDF from a large sample of such realizations.

Let $\theta$ refer to a particular {\em realization} of the integral in  Eq.~(\ref{eq:kappa3b}) and denote the resulting convergence by $\kappa(\theta,z_s)$.  Because of the finite size of the halos and filaments, only a finite number $N_\theta$ of objects intercept the geodesic and contribute to the sum. Moreover, because all objects are small compared to a typical comoving distance to the source $r_{s} \gg R,L$, the function $G$ is, to a good approximation, a constant $G(r,r_s) \approx G(r_j,r_s)$ for each individual object. Similarly, one can assume that $\varphi(x,t) \approx \varphi(x,t_j)$ with $t_j = t(r_j)$, and so one finds:
\begin{equation}
\kappa_{HL} (\theta, z_s) 
\simeq \sum_{j=1}^{N_\theta}G(r_j,r_s) \,\Sigma_{j}(t_j) \,,
\label{eq:realization1}
\end{equation}
where $\Sigma_j$ is the surface mass density of the $j$-th intercepting object in the realization.
Next divide the geodesic into $N_S$ subintervals with centers at $\bar{r}_i$ and with (possibly variable) length $\Delta r_i \ll r_{s}$, such that $G(r,r_s)\approx G(\bar r_i,r_s) \equiv G_i(r_s) =\Delta r_{i}^{-1} \int_{\Delta r_{i}} G(r,r_{s}) dr$ holds within each interval. In this way we can write 
\begin{equation}
\kappa_{HL} (\theta, z_s) \simeq \sum_{i=1}^{N_S}\sum_{{j_i}=1}^{N_i} G_i(r_s) \Sigma_{j_i}(t_i) \,,
\label{eq:realization2}
\end{equation}
where $j_i$ labels all objects encountered by the geodesic within the comoving length bin $\Delta r_i$, such that $\sum_{i=1}^{N_S} N_i = N_\theta$. At this level Eq.~(\ref{eq:realization2}) still consists of a sum over all {\em objects} along a given path. Next we categorize these objects into different classes depending on the parameters that define the surface density $\Sigma_{j_i}$. For spherical halos of a given $\varphi$-profile $\Sigma$ depends only on the mass of the halo and the impact parameter, and for filaments the relevant parameters are the angle between the main filament axis and the geodesic and the impact parameter in the cylindrical radius. Now assume all these variables are discretized into finite-length 
bins.\footnote{All other parameters, such as filament length, radius and mass are kept {\em fixed}. In a more accurate modeling the fixed parameters (for example the filament mass) could be let vary, in which case they should also be discretized.} 
The precise way of the discretization will be discussed in detail in Section \ref{mbin}, but for now we simply let a generic index $u$ label the set of independent parameter cells created by the binning. If we now let $k^\theta_{iu}$ denote the number of objects that fall into the cell labeled by indices $i$ and $u$ in the realization $\theta$, we can replace the sum over the individual objects in Eq.~(\ref{eq:realization2}) by a weighted sum over the discrete parameter cells:
\begin{eqnarray}
\kappa_{HL} (\theta, z_s) &\simeq &
\sum_{i=1}^{N_S}  G_i(r_s) \sum_{u=1}^{N_{C}} k^\theta_{iu} \, \Sigma_{iu}(t_i) 
\nonumber \\
&\equiv& \sum_{i=1}^{N_S} \sum_{u=1}^{N_{C}} k^\theta_{iu} \, \kappa_{1iu}(z_s) \,,
\label{eq:realization3}
\end{eqnarray}
where $N_{C}$ is the total number of independent parameter 
cells\footnote{The binning of internal variables could depend on $r$, and so the indexing $u$ and the number of different parameter cells $N_C$ could in fact depend on $i$, but we suppress this notation for simplicity.}
used in the binning, $\sum_{i=1}^{N_S} \sum_{u=1}^{N_{C}} k^\theta_{iu} = N_\theta$ and $\kappa_{1iu}$ is the convergence due to one object in the bin $i u$:
\begin{equation}
\kappa_{1iu}(z_s)\equiv  G_i(r_s) \, \Sigma_{iu}(t_i) \,.
\end{equation}
Note that the quantity $\kappa_{1iu}$ is a function of the distance and internal variables, it is universal for arbitrary realizations: all the information specific to a particular realization $\theta$ is contained in the set of integers~$\{k^\theta_{iu}\}$.

Equation (\ref{eq:realization3}) is the starting point of the sGL analysis,  because it can be easily turned into a probabilistic quantity. Indeed, instead of thinking of realizations $\theta$ along arbitrary lines of sight through a pre-created model universe, we can define {\em a statistical distribution of convergences} through Eq.~(\ref{eq:realization3}). Indeed, we have shown in Ref.~\cite{Kainulainen:2009dw} that, for random initial and final points of the geodesics, the integers $k^\theta_{iu}$ are distributed as Poisson random variables:
\begin{equation}
P_{k_{iu}} = \frac{(\Delta N_{iu})^{k_{iu}}}{k_{iu}!} \, e^{-\Delta N_{iu}} ,
\label{eq:poisson-i}
\end{equation}
where the parameter $\Delta N_{iu}$ is the {\em expected} number of objects in the bin volume $\Delta V_{iu}$:
\begin{equation} 
\Delta N_{iu} = \Delta n_{iu} \, \Delta V_{iu} = \Delta n_{iu} \, \Delta r_i\, \Delta A_{iu} \, .
\label{deltan}
\end{equation}
Here $\Delta n_{iu}$ is the comoving density of objects corresponding to the parameters in the bin $iu$ and $\Delta A_{iu}$ is the corresponding cross sectional area of these objects in co-moving units. We shall specify these quantities precisely in Section \ref{mbin}. Physically, the statistical distribution of convergences is equivalent to our original set of realizations $\theta$ averaged over the position of the observer. This is a welcome feature because the statistical model explicitly incorporates the Copernican Principle. The original realization $\theta$ has now been replaced by {\em a configuration} of random integers $\{k_{iu}\}$ and the  convergence equation Eq.~(\ref{eq:realization3}) by a statistical convergence:
\begin{equation}
\kappa_{HL} (\{k_{iu}\},z_s) = \sum_{i u} k_{iu} \, \kappa_{1iu}(z_s)  \,,
\label{eq:kappaMfinal}
\end{equation}
where the probability for a particular configuration to occur is just
\begin{equation}
P_{\{k_{iu}\}} = \prod_{i=1}^{N_S}  \prod_{u=1}^{N_C}  P_{k_{iu}}  \,.
\label{eq:Prob1}
\end{equation}
It is easy to show that this probability distribution is correctly normalized. Moreover, including the convergence $\kappa_{UE}$ due to the uniformly distributed matter and the empty space (see Eq.~(\ref{kem}) and below), one can show that the total convergence corresponding to the configuration $\{k_{iu}\}$  becomes~\cite{Kainulainen:2009dw}:
\begin{equation}
\kappa(\{k_{iu}\},z_s) = \sum_{iu} \,\kappa_{1iu}(z_s)  \Big(k_{iu}-\Delta N_{iu} \Big) \,.
\label{eq:fullkappa}
\end{equation}
For a configuration without any inhomogeneities we correctly recover $\kappa(\{k_{iu}=0\},z_s)= \kappa_{UE}(z_s)$, see Section \ref{mbin}. Moreover, 
Eq.~(\ref{eq:fullkappa}) shows explicitly that the {\em expected} convergence vanishes consistently with the photon conservation in weak lensing, because $\langle k_{iu}\rangle =\Delta N_{iu}$ for a random variable following Eq.~(\ref{eq:poisson-i}).  The final convergence PDF can now be formally written as
\begin{equation}
P_{\rm wl}(\kappa, z_s) 
= {\rm lim}_{\Delta \rightarrow 0} \frac{1}{\Delta}
\int_{\kappa-\frac{\Delta}{2}}^{\kappa+\frac{\Delta}{2}} {\rm d}\kappa'
\hat P_{\rm wl}(\kappa', z_s)  \,,
\label{eq:PDF2a}
\end{equation}
where
\begin{equation}
\hat P_{\rm wl}(\kappa, z_s) \equiv \sum_{\{k_{iu}\}} P_{\{k_{iu}\}} 
                 \delta(\kappa - \kappa(\{k_{iu}\},z_s)) \,,
\label{eq:PDF1}
\end{equation}
is a discrete probability distribution.
Note that the most likely configuration which maximises $P_{\{k_{iu}\}}$ corresponds to the mode of the Poisson distribution, which is the floor of its parameter: $k_{iu} \rightarrow \lfloor \Delta N_{iu}\rfloor$.
Moreover, for a large $\Delta N_{iu}$ the Poisson distribution approximates a gaussian with mean and variance equal to $\Delta N_{iu}$ and the most likely configuration therefore approaches the mean: $k_{iu} \rightarrow \Delta N_{iu}$.
When this is the case the mode of the lensing PDF vanishes even for a single observation and the PDF tends to a gaussian.
One can similarly create the PDF in the shift in the distance modulus $\Delta m$ or in the magnification $\mu$, using Eq.~(\ref{eq:mukappa}) to compute these quantities for a given configuration\footnote{This is true assuming that nonlinear effects can be neglected, {\em i.e.}, in the weak lensing limit.} and then replacing $\kappa$ by the desired quantity in equations (\ref{eq:PDF2a}-\ref{eq:PDF1}).

In practice, equations (\ref{eq:PDF2a}-\ref{eq:PDF1}) are not useful for a direct evaluation of the PDF. Indeed, the configuration spaces are infinite and a direct evaluation of the sum of configurations would not be feasible even for relatively roughly discretized systems. Instead, we compute an approximation for $P_{\rm wl}$ by statistically creating a large set of random configurations $\{k_{iu}\}$ from the probability distribution $P_{\{k_{iu}\}}$, and by forming a discrete normalized histogram out from the corresponding set of convergences (or distance moduli or magnifications). That is, we define a histogram PDF using the function:
\begin{equation}
P_{\rm wl}(\bar \kappa_i, z_s)\Delta_{i}  \approx \frac{N_{{\rm sim},i}}{N_{\rm sim}} \,,
\label{eq:PDF2}
\end{equation}
where $N_{\rm sim}$ is the total number of realizations and $N_{{\rm sim},i}$ is the number of convergences in the sample falling in the bin $\Delta_{i}$ centered around the mean value $\bar \kappa_i$.
It is easy to check that Eq.~(\ref{eq:PDF2}) is correctly normalized to unity.
Moreover, the more configurations one creates in the simulation step, the more fine-detailed and accurate approximation for $P_{\rm wl}(\kappa , z_s)$ can be constructed.\footnote{One could also define a smooth $P_{\rm wl}(\kappa, z_s)$ from the simulated data as a moving average weighted by some window function, but that would not give any advantage in the analysis with respect to using the simple histogram of Eq.~(\ref{eq:PDF2}).}

To summarize, the sGL method consists of two steps: first the internal variables specifying the inhomogeneities are binned, reducing them to a finite number of different object classes labeled by the cell index $u$, for which universal convergence functions $\kappa_{1iu}$ can be computed. Second the occupation numbers $k_{iu}$ of these objects within a given co-moving distance bin are shown to be Poisson distributed with parameter $\Delta N_{iu}$. The lensing PDF is then computed statistically, binning the convergence distribution obtained from a large set of random configurations $\{k_{iu}\}$ on a given geodesic.
Computing the lensing PDF this way is very efficient. Finding the PDF for a given model universe using the {\tt turboGL} package~\cite{turboGL} typically takes a few seconds in an ordinary desktop computer. See the Appendix \ref{perf} for more details about its performance.

Let us point out that Eq.~(\ref{eq:PDF2}) treats all objects on the same footing. That is, all object families are independently randomly distributed throughout the space. Confining the halos to within filaments will require slight modifications which we will explain in the section~\ref{confi}. For now, we shall study the effects that may cause the actual observed PDF deviate from the fundamental weak lensing PDF.

\subsection{Observable PDF}
\label{sec:observations}

The lensing PDF (\ref{eq:PDF2}) is the fundamental quantity inherent to a given background model and a set of inhomogeneities. However, in reality there are interferences that prevent us from observing $P_{\rm wl}$ directly. Firstly, the intrinsic magnitudes of the sources are not known with arbitrary accuracy, and so the {\em observed} lensing PDF is at best a convolution:
\begin{equation}
P_{1}(\Delta m,z_s) = \int {\rm d} y  \, P_{\rm wl}(y,z_s) \, P_{\rm in}(\Delta m - y) \,,
\label{eq:obsPDF1}
\end{equation}
where $P_{\rm in}$ describes the source magnitude dispersion for an imperfect standard candle. Clearly the fundamental lensing PDF is recovered in the limit that $P_{\rm in}$ becomes a delta function. If there were no other sources of errors, the distribution $P_1$ could be used to contrast observations against different cosmological background models, sets of inhomogeneities and the models for the source magnitude distributions. However, there can be several different types of selection effects caused by matter between the source and the observer or by the search strategies, that may bias the observed PDF from the one predicted by Eq.~(\ref{eq:obsPDF1}). One of the strengths of the sGL method is that it can be easily extended to model the effects of a wide range of selection biases on the observable PDF.

\subsubsection*{Observational Biases}
\label{sbs}

Let us begin with a simple class of biases that might affect the observed distance modulus beyond the weak lensing correction; an obvious example would be an error made in the estimate of reddening. What makes such correction nontrivial, is that one would expect the error to be {\em correlated} with the types of mass concentrations encountered by the photon beam on its way from the source to the observer. The effects should thus be different along different lines of sight, affecting mostly those directions where light has to travel through volumes with higher mass densities. In the sGL method such correlated biases are very easily modelled, as one needs merely to replace 
\begin{eqnarray}
\Delta m &\rightarrow& 
        \Delta m(\{k_{iu}\},z_s) + \delta m(\{k_{iu}\},z_s)
\nonumber \\
&\equiv& \Delta m_{\rm eff}(\{k_{iu}\},z_s) \,,
\label{eq:deltameff}
\end{eqnarray}
where the first part $\Delta m(\{k_{iu}\},z_s)$ is the usual weak lensing correction and the second part the object-specific correction. Indeed, the bias function $\delta m(\{k_{iu}\},z_s)$ can be correlated with the redshifts $z_s$ and $z_i$, and with all the internal characteristics of the objects, such as the halo and filament masses and impact parameters. A quantitative analysis of the possible magnitude of the function $\delta m$ would require detailed astrophysical information and it is beyond the scope of this paper. However, once properly modelled, the sGL allows a very easy way of evaluating the outcome of biases on the observations.

Other types of {\em selection} biases might not directly affect the magnitudes, but rather the relative likelihood of observing particular events, in a way that is correlated with the mass density. This includes, for example, all sources leading to obscuration of the light beam, either alone or in combination with restrictions coming from imperfect search efficiencies and strategies. For example selection by extinction effects or by outlier rejection mainly relate to high-magnification events which are clearly correlated with having high intervening mass concentrations (halos with large $M$ and/or small $b$) along the light geodesic. Similarly, short duration events might be missed by search telescopes or be rejected from the data due to poor quality of the light curve, for example in cases where a supernova is not separable from the image of a bright foreground galaxy. Probability of such events would obviously be correlated with the brightness of the source ($z_s$ and source dependent parameters), with the density and redshift of the intervening matter and of course with the search efficiencies. 

The selection biases can be studied in the sGL framework by introducing a new, object-dependent {\em survival probability function}. That is, we replace
\begin{equation} 
P_{\{k_{iu}\}} \rightarrow P^{\rm eff}_{\{k_{iu}\}} \equiv K \, P^{\rm sur}_{\{k_{iu}\}}P_{\{k_{iu}\}}
\label{eq:survival}
\end{equation}
in the probability distribution for the magnification PDF~(\ref{eq:PDF1}). Here $K$ is a normalization constant that makes sure that the final PDF is normalized to unity. The most generic form of the survival probability function is
\begin{equation}
P^{\rm sur}_{\{k_{iu}\}} = \prod_{iu} \left( P^{\rm sur}_{iu} \right)^{k_{iu}} \,.
\label{eq:Piu}
\end{equation}
This allows the survival function to depend on the arbitrary local properties along the photon geodesic. The physical interpretation of $P^{\rm sur}_{iu}$ is of course clear: it gives the relative probability that an event whose signal intercepts an object with characteristics given by internal parameters $u$ at redshift $z_i$ would make it to the accepted observational sample.
Given the form (\ref{eq:Piu}) the proper normalization is easily seen to be:
\begin{equation} 
P^{\rm eff}_{\{k_{iu}\}} = \prod_{iu} 
   \frac{(\Delta N^{\rm eff}_{iu})^{k_{iu}}}{k_{iu}!} 
         e^{-\Delta N^{\rm eff}_{iu}} \,,
\label{eq:survival2}
\end{equation}
where
\begin{equation}
\Delta N^{\rm eff}_{iu} = P^{\rm sur}_{iu} \Delta N_{iu} \,.
\label{eq:deltaNeff}
\end{equation} 
The correct expression for the convergence is still given by Eq.~(\ref{eq:fullkappa}), with the important difference that the random integers $k_{iu}$ used to evaluate Eq.~(\ref{eq:PDF2}) are now drawn from the Poisson distribution of Eq.~(\ref{eq:survival2}) with parameter the effective expected number of objects $\Delta N^{\rm eff}_{iu}$.

As mentioned above, modeling $P^{\rm sur}_{iu}$ quantitatively would need detailed input of the astrophysical properties of the intervening matter distributions and of the observational apparatus, which is beyond the scope of this work.
See Ref~\cite{Kainulainen:2009dw} for an analytical example giving $\langle \kappa (z_s) \rangle =  (1 - \alpha) \, \kappa_E$, where $\alpha$ is the effective filling factor of a partially-filled beam~\cite{Kantowski:1969} (see \cite{Mattsson:2007tj} for an extension of the partially-filled beam formalism).
We will discuss an illustrative numerical example in Section \ref{results} where we will consider as survival probability a simple step-function in the impact parameter.

To summarise, let us note that it is easy to combine both types of observational biases discussed in this section. The true observational PDF can be defined as
\begin{equation}
P_{\rm obs}(\Delta m,z_s) = \int {\rm d} y  \, P^{\rm eff}_{\rm wl}(y,z_s) \, P_{\rm in}(\Delta m - y) \,,
\label{eq:obsPDF2}
\end{equation}
where $P_{\rm in}$ again is the intrinsic source magnitude dispersion and the effective lensing probability distribution function $P^{\rm eff}_{\rm wl}(y,z_s)$ is computed from Eq.~(\ref{eq:PDF2}) using the effective probability distribution of Eq.~(\ref{eq:survival2}), and computing the effective magnitudes for  configurations from equations (\ref{eq:deltameff}) and (\ref{eq:mukappa}).
This setting clearly allows modeling a very complex structure of observational biases. 

Finally note that the generic trend among all the effects discussed above is to suppress events with higher magnifications. It could then be that a non-negligible overestimate of the observed magnitudes follows from neglecting the selection biases, even if the individual effects were relatively small. Moreover, while the present data cannot yet constrain the lensing PDF itself, it can put bounds on its variance. Since the observational biases are suppressing the high magnification tail, they give $P_{\rm obs}$ a smaller effective variance which may open the possibility for a comparison with the experiments. We will further discuss these topics in Section \ref{results}.

\subsection{Binned Data Samples}
\label{binnedsample}

We conclude this section by deriving an effective distribution $P_{N_O}$ for a data sample $\{ N_O \}$, where the set $\{ N_O \}$ refers to a given binning of the original data. $P_{N_O}$ could be computed directly from the fundamental PDF~\cite{Kainulainen:2009dw}, but it is much faster to create it directly during the initial simulation, bypassing the calculation of the fundamental $P_{\rm wl}$ altogether. Indeed, if we label the configurations within a given sample of observations by $s$, the mean convergence after $N_O$ observations is:
\begin{eqnarray} 
\kappa_{N_O}(\{k_{iu}\})&=& {1 \over N_{O}} \sum_{s=1}^{N_{O}} \kappa(\{k_{iu}\}_{s}) \label{equation:kappa3} \\
&=&\sum_{iu} \kappa_{1iu} \left({ \sum_{s=1}^{N_{O}}  k_{iu, s} \over N_{O}} -\Delta N_{iu} \right) \nonumber \\
&=& \sum_{iu}   \kappa_{1iu} \left( {k_{iu, N_O} \over N_{O}} -\Delta N_{iu} \right) \, , \nonumber
\end{eqnarray}
where in the second line we have used the fact that the quantities $\kappa_{1iu}$ are independent of $s$ and in the last line we have used the property that the independent Poisson variables $k_{iu, s}$ (with the same weight) sum exactly (in a statistical sense) into the Poisson variable $k_{iu, N_O}$ of parameter given by the sum of the individual parameters, which is $N_O \, \Delta N_{iu}$.
$P_{N_O}$ is then given by an expression similar to Eqs.~(\ref{eq:PDF2a}-\ref{eq:PDF1}) or Eq.~(\ref{eq:PDF2}).

Note, however, that including the selection effects (see Section \ref{sbs}) in general does not commute with taking an average over the observations. In other words, if the $N_{O}$ measurements are correlated, we cannot use Eq.~(\ref{equation:kappa3}), but we have to start from the fundamental PDF. On the other hand, Eq.~(\ref{equation:kappa3}) displays explicitly the effect of the size of the data sample: even if $\kappa$ had a skewed PDF and a nonzero mode for $N_{O}=1$, for large $N_{O}$ the distribution approaches a gaussian and eventually converges to a $\delta$-function with a zero convergence, as it is clear from the properties of Poisson variables.

\subsection{Integral Formulas for Expected Value and Variance of the Lensing Convergence}
\label{analvar}

From the general expression Eq.~(\ref{equation:kappa3}), including a nontrivial survival probability, it follows that the expected value and variance of the convergence are:
\begin{equation}
\langle \kappa \rangle 
= \sum_{iu} \kappa_{1iu}  \, (P^{\rm sur}_{iu} - 1)\Delta N_{iu} \,,
\label{averagekappa}
\end{equation} 
and
\begin{equation}
\sigma^{2}_{\kappa}=  {1 \over N_{O}}  \sum_{iu} \,\kappa_{1iu}^{2} \, P^{\rm sur}_{iu} \, \Delta N_{iu} \,,
\label{varkappa}
\end{equation}
where we assumed that the survival probability for a light ray going through the uniformly distributed matter density $\rho_{MU}$ equals unity and that the random variables $k_{iu}$ are uncorrelated.
Eqs.~(\ref{averagekappa}-\ref{varkappa}) can then be put back in integral form giving the following {\em exact} results:
\begin{eqnarray} \label{averagekappax}
 \langle \kappa \rangle &=&    \int_{0}^{r_{s}} dr \, G(r,r_{s}) \int_{M_{\rm cut}}^{\infty} dn(M,z(r)) \times \\
& \times& \int_{0}^{R(M,z(r))}dA(b,M) \,  (P_{\rm sur}  -1) \,  \Sigma(b,M,z(r)))  \,,  \nonumber
\end{eqnarray}
and
\begin{eqnarray} \label{varkappax}
\sigma^{2}_{\kappa}&=&   {1 \over N_{O}} \int_{0}^{r_{s}} dr \, G^{2}(r,r_{s}) \int_{M_{\rm cut}}^{\infty} dn (M,z(r)) \times \\
&& \times \int_{0}^{R(M,z(r))}dA(b,M)  \, P_{\rm sur}   \, \Sigma^{2}(b,M,z(r)))  \,, \nonumber
\end{eqnarray}
where the integral limits for the last two integrals are implicitly defined and the explicit form of surface area element $dA$ depends on the geometrical properties of the object of mass~$M$. For example, for spherical halos it is $dA=2 \pi b \, db$.
We again stress that the quantity $P_{\rm sur}=P_{\rm sur}(b,M,z,z_{s})$ is a generic function that describes the probability that a light ray is observed when it hits a halo of mass $M$ at the redshift $z$ and impact parameter $b$ for a source at redshift~$z_{s}$.

Eqs.~(\ref{averagekappax}-\ref{varkappax}) are a simple and direct prediction of the sGL method and allow to draw some general considerations.
First, if $P_{\rm sur}=1$, the expected value of the convergence is correctly zero, showing the ``benevolent'' nature of weak lensing corrections for unbiased observations. If however the survival probability is not trivial, selection biases persist even in very large datasets so that the average convergence over a large number of observations approaches a nonvanishing value.
Second, Eq.~(\ref{varkappax}) is a product of positive quantities and so {\em a nontrivial survival probability  always reduces the observed variance.}

We would like to stress that it is easy to numerically evaluate the integrals of Eqs.~(\ref{averagekappax}-\ref{varkappax}), and their predictions can be straightforwardly implemented in $\chi^{2}$ analyses based on gaussian likelihoods, which may be a reasonable approximation if the observable PDF is not strongly skewed.

\section{Discretizing the model parameters} 
\label{mbin}

The key role played by the discretization of the model parameters in the sGL method is now quite obvious. Of course, the final lensing PDF should not depend on the discretization. Indeed, as the numerical value of a Riemann integral does not depend on the binning chosen if the integrand does not significantly vary within the bins, the resulting PDF will not depend on the specific binning if the cosmological functions $\kappa_{1iu}$ are approximately constant within the bin cells $iu$. Once this condition is well met a further refinement of binning does not increase the accuracy. This statement can be proved formally as follows: imagine that we have a valid binning prescription $\mathcal{A}$ and we want to move to a new prescription $\mathcal{B}$ which is $N$ times finer. Because ${\mathcal A}$ is valid, the convergences $\kappa^{\mathcal B}_{1iu}$ are approximately constant within each $\mathcal{A}$-bin. The relative Poisson variables can then be resummed as done in Eq.~(\ref{equation:kappa3}), and we obtain the same formal expression in terms of ${\mathcal A}$-bins we started from.

This theorem also explains why we do not need to bin any variable in which the surface mass density $\Sigma$ of a given object is constant, such as the rotational angle along the center of a spherical halo in the lens plane.
As the convergences $\kappa_{1iu}$ are constants within angular bins, their contributions can be resummed to give a model without any angular binning. That is, the fundamental halo objects are actually finite-width rings in the convergence plane. 

The first quantity to be binned is the co-moving distance to the source. Here the basic quantity that needs to be accurately modelled is the function $G(r,r_s)$, which gives the optical weight of a given convergence plane.
This is a rather smooth function with a shallow peak, and only slightly dependent on $r_s$ when expressed in the scaled variable $r/r_s$. Simple linear bins $\bar{r}_i$ have been found to give accurate results, with only $10-15$ redshift slices.
This turns out to be the typical number of bins needed for an accurate modeling of most of the internal variables as well.

\subsection{Discretizing the Halo Parameters}
\label{binning}

\begin{figure}
\begin{center}
\includegraphics[width=8.0 cm]{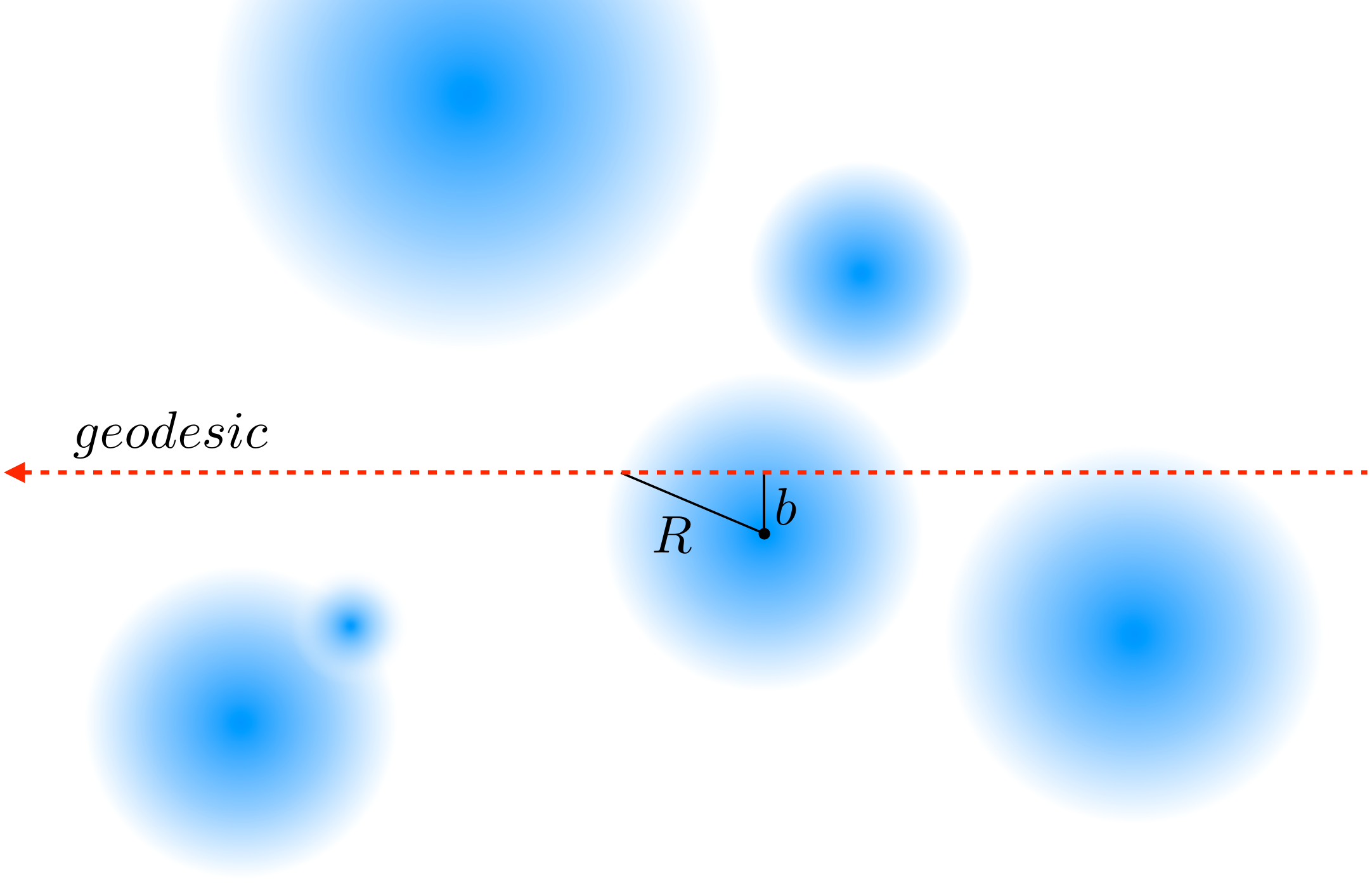}
\caption{Shown is a comoving segment of a photon geodesic intercepting a number of halos represented by shaded discs.}
\label{sketch}
\end{center}
\end{figure}

We begin by discretizing the halo parameters, which are the mass and the impact parameter in the lens plane. Binning the halo masses involves two main considerations.
First, the halo mass function only carries information about high-contrast virialized halos within a specific mass range (see Eq.~(\ref{jenk})).
Moreover, as explained in Section \ref{mati}, only very large mass concentrations play a significant role in our weak lensing analysis.
Consequently, we introduced the lower cut $M_{\rm cut}$ to the mass distribution so that only mass concentrations with $M>M_{\rm cut}$ are treated as halos, while the remaining mass has been divided into low contrast objects and uniform matter density (see Eq.~(\ref{eq:sumrule})).
Second, the fact that the concentration parameter of Eq.~(\ref{duffy}) follows a power law suggests the use of a logarithmic binning in mass. The mass function drops quickly at very large masses, so that we can combine all masses above a certain limit to a single bin.
We found it reasonable to define at any redshift\footnote{A more efficient binning may be obtained by adopting redshift-dependent mass bins ${M}_{n}(z)$.
One could, for example, use bins such that $\Delta f_{LU}$ has always the present-day value. See Section \ref{mati} for a discussion about $\Delta f_{LU}$.} the cell boundaries by ${M}_{n}= 10^{n} \, h^{-1} M_{\odot}$ and the cell centers by $\bar M_{n}= 10^{n+1/2} \, h^{-1} M_{\odot}$, with $M_{\rm cut}=M_{10}$.
Because $f(M>M_{16},z) \simeq 0$ to a very good approximation, it is sufficient to bin with an integer $n$ between 10 and 15:
\begin{eqnarray} 
M_{10}< &M& \le M_{11} \quad \textrm{``companion galaxies",} \nonumber \\
M_{11}< &M& \le M_{12} \quad \textrm{``spiral galaxies",} \nonumber \\
M_{12}< &M& \le M_{13} \quad \textrm{``elliptical galaxies",} \nonumber \\
M_{13}< &M& \le M_{14} \quad \textrm{``groups",} \nonumber \\
M_{14}< &M& \le M_{15} \quad \textrm{``clusters",} \nonumber \\
M_{15}< &M& \le \infty  \qquad \textrm{``superclusters".} 
\label{mbins}
\end{eqnarray}
The fraction of total mass in a bin is then
\begin{equation}
\Delta f_{n}= \int_{M_{n}}^{M_{n+1}} f \, {\rm d}M  \, ,
\end{equation}
which correctly gives the total halo mass fraction (see Eq.~(\ref{eq:def_of_fH})) when summed over all the halo bins:
\begin{equation}
\sum_{n} \Delta f_{n} = \Delta f_{H} \, .
\end{equation}
Finally, the binned comoving halo densities are
\begin{equation} \label{nbins}
\, \Delta n_{n} \equiv \frac{\rho_{MC}}{\bar M_{n}}  \, \Delta f_{n} \, ,
\end{equation}
which also allows to relate the mass $\bar M_n$ to the interhalo scale $\lambda_{n} \equiv \Delta n_{n}^{-1/3}$.

The halo impact parameter $b_{h}$ in the lens plane is restricted to the range $b_{h} \in [0,R]$, where $R$ is the comoving halo radius.
It should be noted that through Eq.~(\ref{mdel}) the radius depends both on the halo mass and the redshift.
See Fig.~\ref{sketch} for an illustrative sketch.
It is efficient to discretize $b$ into bins of constant integrated surface density (constant equal mass):
\begin{equation}
 \int_{b^{h}_{i n m}}^{b^{h}_{i n m+1}} {\rm d}b \, 2\pi b \, \Sigma_h(b,\bar M_n,t_{i}) \equiv \frac{\bar M_n}{N^{h}_B}   \,,
\label{binbounds}
\end{equation}
where $N^h_B$ is the number of $b$-bins used and the surface density is given by (see Fig.~\ref{sketch}):
\begin{equation}
\Sigma_h(b_{h}, M, t) = M \int_{b_{h}}^{R} \frac{2 \,   x \, {\rm d}x}{\sqrt{x^2-b_{h}^2}} \, \varphi_{h}(x,t) \,.
\end{equation}

After the redshift and the halo mass is specified, the bin boundaries $b^{h}_{inm}$ and the weighted centers of gravity $\bar b^h_{inm}$ can be computed using Eq.~(\ref{binbounds}).
After these are defined, the corresponding area functions $\Delta A^h_{inm}=\pi \, b^{{\ssh} \, 2}_{i n m+1} - \pi \, b^{{\ssh} \, 2}_{i n m}$, needed for the Poisson parameters $\Delta N^h_{inm}$ and the binned surface densities $\Sigma^h_{inm} = \Sigma_h(\bar b^h_{inm},\bar M_n,t_i) \equiv \bar M_n / ( N^{h}_B \Delta A^h_{inm} )$, can be computed.

Overall, our halo model is described by the redshift index and two internal indices for the discretized mass and impact parameter. In terms of the generic parameter $u$ we can formally express this as: $\{h, inm\}\in \{iu\} $, where the label $h$ refers to the ``halo" family of the inhomogeneities.

\subsection{Discretizing the Filament Parameters} 
\label{moddof}

\begin{figure}
\begin{center}
\includegraphics[width=8.0 cm]{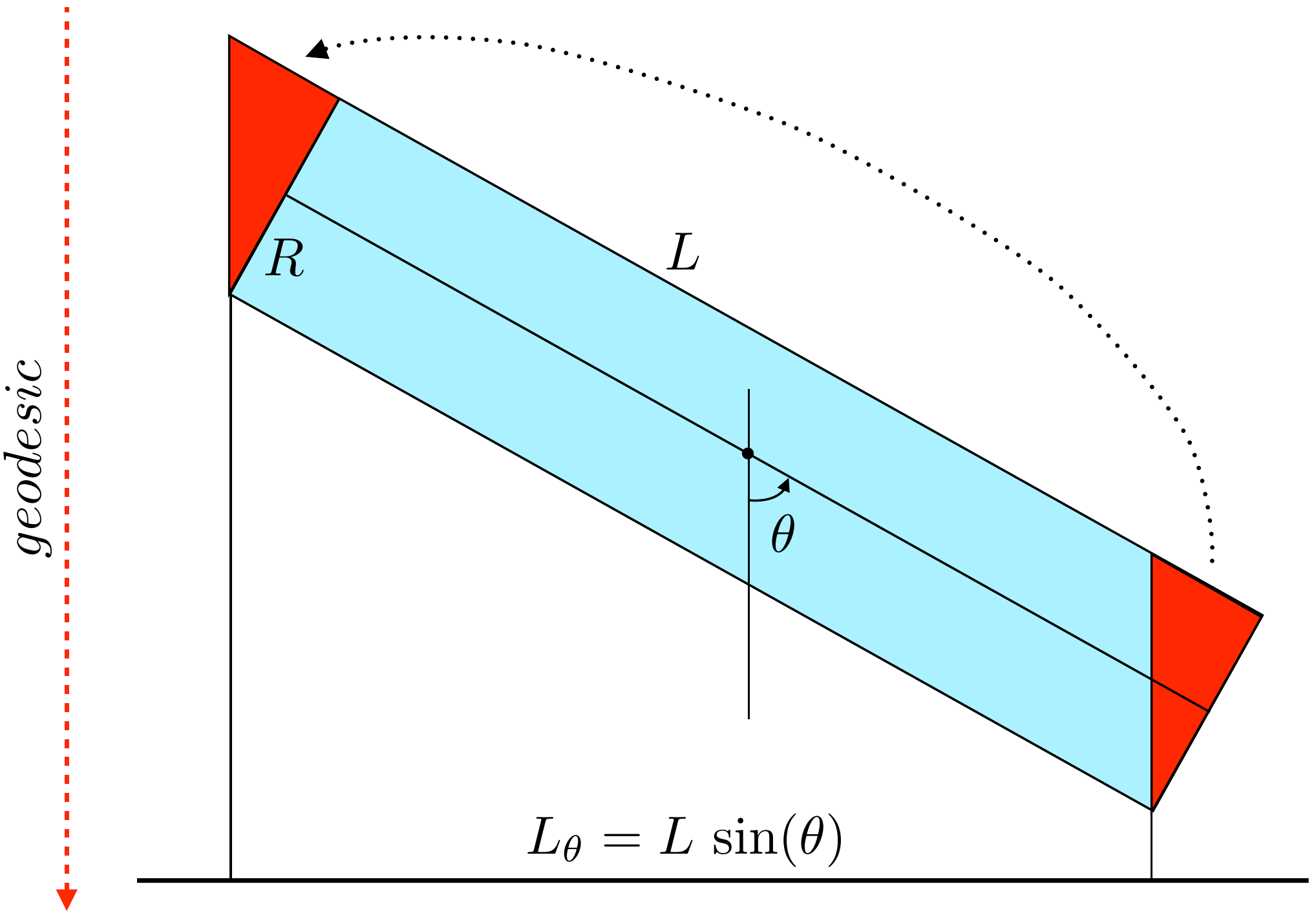}
\caption{This illustration shows how we project the cylindrical filaments on the plane perpendicular to the geodesic.
The triangular section on the right is moved to the left end of the filament so that the projected profile is independent of $l_{\theta}$.}
\label{cylinder2}
\end{center}
\end{figure}

\begin{figure}
\begin{center}
\includegraphics[width=8.0 cm]{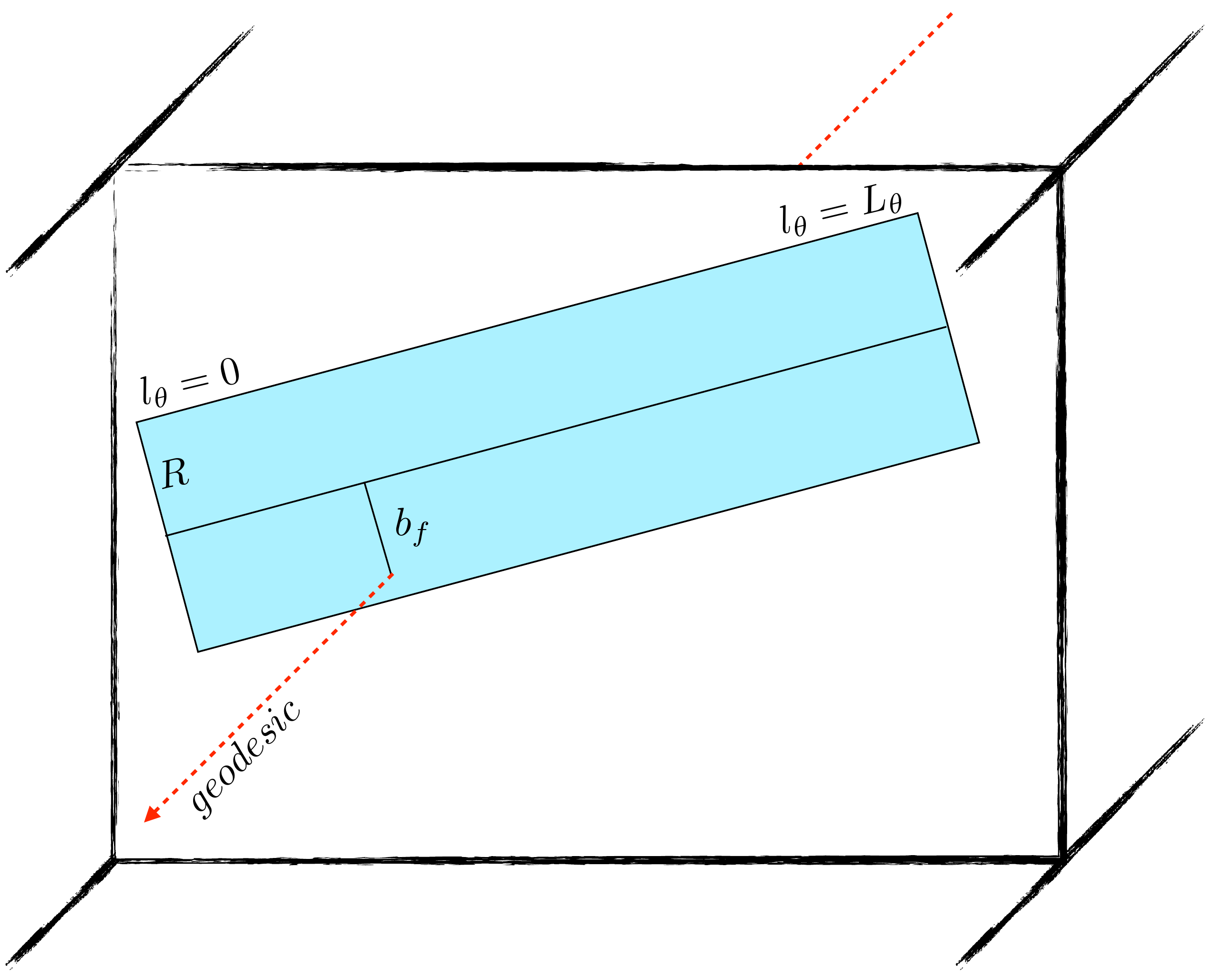}
\caption{Sketch of a cylindrical filament projected on a plane perpendicular to the geodesic. The effective length of the cylinder is $L_\theta$ of Eq.~(\ref{ltt}) as shown in Fig.~\ref{cylinder2}.}
\label{cylinder1}
\end{center}
\end{figure}

The position and the orientation of a generic object in the three dimensional space is described by three position and three orientation degrees of freedom. Since our filaments objects are confined to within a given $\Delta r_{i}$ slice, they generically have only five degrees of freedom. Special symmetries of a sphere reduced the number of relevant parameters for halos to a sigle one, the impact parameter $b_{h}$. For filaments the situation is slightly more complicated because a cylinder is invariant only for rotations around one symmetry axis. This leaves us with four degrees of freedom.
The relevant parameters, however, are reduced by the symmetries to only two: the angle $\theta$ of the filament main axes with respect to the geodesic and the impact parameter $b_{f}$. For illustration see Figs.~\ref{cylinder2}-\ref{cylinder1}.
This can be understood by the fact that all the possible filament configurations with the same $\theta$ and $b_{f}$ have the same surface density $\Sigma_f$ and so can be resummed using the theorem given above.
In particular, the projected surface density of our filaments does not depend on the coordinate $l_{\theta}$ along the projected main axes\footnote{In order to have the projected profile independent of $l_\theta$ we ``move" the triangular section as shown in Fig.~\ref{cylinder2}. Strictly speaking our filaments do not have a cylindrical symmetry, but a $\theta$-dependent ``tilted" rotation symmetry, such that the projected surface densities follow equation (\ref{gammat0}) for all $l_\theta$. This approximation should be good for long filaments with $L \gg R$.}
allowing to resum the $l_{\theta}$-bins into a single effective bin of the total $\theta$-dependent length of the projection:
\begin{equation} \label{ltt}
L_{\theta} = L \, \sin \theta \,.
\end{equation}
The angle $\theta$ is physically important because objects seen with a smaller $\theta$ have a smaller cross section and a higher surface density. Accounting for this geometrical effect leads to a more strongly skewed lensing PDF.

The expression for the surface density is given by
\begin{equation} \label{gammat0}
\Sigma_f (b_f,\theta ,t) = {M_{f} \over \sin \theta} \int_{b_f}^{R} 
\frac{2 \, x \, {\rm d}x}{{\sqrt{x^2-b_{\ssf}^2}}} \, \varphi_{f}(x,t) \,.
\end{equation} 
For the uniform filament profile this gives simply $\Sigma_f = 2 \, \bar \rho_{f\theta}   \sqrt{R^2-b_{\ssf}^2}$, where $\bar \rho_{f\theta} \equiv M_{\ssf}/(\pi R^2 L_\theta)$. For the gaussian profile (\ref{filagau}) one instead finds
\begin{equation} 
\Sigma_f  
= 0.997^{-1} \bar \rho_{f\theta} \, R \, \sqrt{\frac{9\pi}{2}}  \exp[- b_f^2 / 2(R/3)^2]   \,.
\label{gammat}
\end{equation}

The angle $\theta$ can be restricted by symmetry to the interval $\theta \in [0,\pi/2]$, which we divide into $N_T$ uniform length bins around centers $\bar \theta_t = \pi(t-1)/(2N_T)$.
The impact parameter is restricted to $ b_f  \in [0,R]$, and it will be discretized into $N^{f}_B$ bins using the equal mass criteria, as in the halo case defined in Eq.~(\ref{binbounds}). That is, our filaments will be represented by a family of ``bars" of different surface densities $\Sigma^{\ssf}_{itm}$ and length $L_t \equiv L \sin \bar\theta_t$.
The surface area of these bars to be used in the Poisson parameters is $\Delta A^{\ssf}_{itm} =2 \, L_t \, \Delta b^{\ssf}_{itm}$.

Overall, our filament model is described by the redshift index, and two internal indices for the discretized angle and impact parameter. In terms of the generic parameter $u$ we can again formally express this as: $\{f,itm\}\in \{iu\} $, where the family label $f$ refers to the ``filament" family of the inhomogeneity. Of course adding, {\em e.g.}, a mass distribution for filaments and/or an $l$-dependent density profile would increase the necessary number of indices in the discrete model.

\section{Confining Halos to filaments}
\label{confi}

Until now we have implicitly assumed that the halo and the filament families are independent. We will now show how a more realistic model with halos confined to the filaments can be set up in the sGL approach. The central observation is that the exact positions of halos and filaments within the co-moving distance bins are irrelevant in Eq.~(\ref{eq:fullkappa}) for the convergence. It then does not matter if we really confine the halos to filaments; for the same effect it is sufficient to merely confine them into the equivalent volume occupied by the filaments in a given bin. 

We can actually do even better than imposing a simple volume confinement. Indeed, it is natural to assume that the halo distribution follows the smooth density profile that defines the filament, which can be accounted for by weighting the volume elements by their reduced density profiles. Because only the projected matter density is relevant for lensing, this weighting is accounted for by the effective surface densities of the objects. We then introduce the effective co-moving thickness for the low density objects as:\footnote{Note that this discussion is again general, and applies to all types of low density objects, not necessarily with cylindrical geometry. To distinguish filaments from halos we use $v$ to refer to generic internal filamentary degrees of freedom.}
\begin{equation}
\Delta r^{\ssf}_{iv} \equiv \frac{1}{\bar \rho_{fi}} \, \Sigma^{\ssf}_{iv} \,,
\end{equation}
where $\bar \rho_{fi} \equiv M_f/V_{fi}$ is the average mass density of the filament with volume $V_{fi}$.
The total effective length of a configuration of filaments $\{k^{\ssf}_{iv}\}$ along the line of sight is then:
\begin{equation}
\Delta r^{\ssf}_{i}(\{k^{\ssf}_{iv}\}) =  \sum_v k^{\ssf}_{iv} \Delta r^{\ssf}_{iv}
                         \equiv  \Delta r_i \, q_{fi} \, .
\label{eq:master1}
\end{equation}
It is easy to show that the statistical average of $\Delta r^{\ssf}_{i}$ over the configuration space is just the expected distance covered by filaments:
\begin{equation}
\langle \Delta r^{\ssf}_{i}\rangle =  \sum_v \Delta N^{\ssf}_{iv} \, \Delta r^{\ssf}_{iv} =  \Delta r_i \, \bar q_{fi} \,,
\label{eq:master2}
\end{equation}
where we defined the comoving average filament volume fraction:
\begin{equation}
\bar q_{fi}= V_{fi} \, \Delta n_{fi}  \,.
\end{equation}

The confining simulation can now be performed in two steps: in the first step a random configuration of filaments is generated and Eq.~(\ref{eq:master1}) is used to compute the effective lengths $\Delta r^{\ssf}_i$ at different co-moving slices. These lengths define then the Poisson parameters for the halos through Eq.~(\ref{deltan}):
\begin{equation} 
\Delta N^{\ssfh}_{iu}(\{k^{\ssf}_{iv}\}) = \Delta n^{\ssfh}_{iu}  \, \Delta r^{\ssf}_i  \, \Delta A^h_{iu}  \,.
\label{deltanhalos}
\end{equation}
Note that the halo densities $\Delta n^{fh}_{iu}$ are corrected to account for the reduced volume available due to confinement:
\begin{equation}
\Delta n_{iu}^{\ssfh} \equiv {\Delta n^h_{iu} \over \bar q_{fi}} \,,
\label{eq:haloweights}
\end{equation}
where $\Delta n^h_{iu}$ are the usual halo densities of Eq.~(\ref{nbins}) when the halos are uniformly distributed through all space. Inserting Eq.~(\ref{eq:haloweights}) back to (\ref{deltanhalos}) and using Eqs.~(\ref{eq:master1}-\ref{eq:master2}) we can rewrite Eq.~(\ref{deltanhalos}) simply as
\begin{equation} 
\Delta N^{\ssfh}_{iu} = {q_{fi} \over \bar q_{fi}} \, \Delta N^h_{iu} 
 \equiv X_{fi} \, \Delta N^h_{iu} \,,
\label{deltanf}
\end{equation}
where $\Delta N^h_{iu}$ are the usual configuration independent Poisson parameters computed without confinement.

Note that the central quantities needed in the evaluation of Eq.~(\ref{deltanf}) are the filamentary weights $\Delta N^{\ssf}_{iv}$ and surface densities $\Sigma^{\ssf}_{iv}$.
With these given, all the information about the specific low contrast objects used is condensed into the single stochastic variable:
\begin{equation}
X_{fi} = \frac{\sum_v k^{\ssf}_{iv} \Sigma^{\ssf}_{iv}} 
              {\sum_v \Delta N^{\ssf}_{iv} \Sigma^{\ssf}_{iv}} \,,
\end{equation}
which has an expectation value of unity and a mode smaller than unity.
The lensing effects due to large-scale structures are thus tied to the probability distribution of $X_{f}$: its added skewness is the effect coming from confining halos within filaments. This opens the possibility to investigate and compare different filamentary geometries by means of the $X_{f}$-PDF.
We will develop these thoughts further in a forthcoming paper \cite{SM3}.

We can further generalize the picture given in this Section by considering different levels of confinement for halos of different masses.
Indeed, because halos of different mass are treated independently in Eq.~(\ref{deltanf}), one could have small halos populate also the voids and massive halos only the filamentary structure.

Let us finally point out that in practical computations the convergence PDF is obtained by creating a large number of independent halo configurations for each ``master" filament configuration. Typically we use simulations with few hundred master configurations with a few hundred halo configurations each.

\subsection{Power Spectrum} 
\label{discu}

In this Section we will discuss the power spectrum of our model universe and the importance of large-scale correlations for the lensing PDF. We start by making a connection between the sGL method and the so-called ``halo model'' (see, for example, \cite{Neyman:1952, Peebles:1974, Scherrer:1991kk, Seljak:2000gq, Ma:2000ik, Peacock:2000qk, Scoccimarro:2000gm} and \cite{Cooray:2002dia} for a review), where inhomogeneities are approximated by a collection of different kinds of halos whose spatial distributions satisfies the linear power spectrum. The idea behind the halo model is that on small scales (large wavenumbers $k$) the statistics of matter correlations are dominated by the internal halo density profiles, while on large scales the halos are assumed to cluster according to linear theory. The two components are then combined together.

The sGL modeling of the inhomogeneities can be thought as a two-step halo model where we first create the random filamentary structures and then place the halos randomly within these structures. Similarly to the halo model, one can then combine the linearly-evolved power spectrum with the nonlinear one coming from the filaments and the halos they contain. In this sense our filamentary structures extend the halo model by introducing correlations in the intermediate scales between the halo substructures and the cosmological scale controlled by the linear power spectrum. In the halo model the power spectrum can be computed analytically, but here the calculation has to be done numerically. We are currently extending the sGL method such that the simulation will produce also the power spectrum~\cite{SM3} in addition to the lensing PDF.
The power spectrum can indeed place useful constraints on the filament parameters which, as remarked in Section \ref{mati}, are not tightly constrained by observations.

While the power spectrum is relevant for understanding the correlations at the largest observable scales, the lensing PDF depends mainly on the much stronger inhomogeneities at smaller scales. This can be understood from Eq.~(\ref{eq:kappa1}) which shows the direct dependence of the lensing convergence on the density contrast. The small lensing variance from the large-scale correlations induced by the linear power spectrum was numerically computed, for example, in~\cite{Bernardeau:1996un}. Moreover, weak lensing and the power spectrum probe somewhat different aspects of the inhomogeneities. The web-like structures of filaments and voids that affect weak lensing are mainly described by higher order correlation terms beyond the power spectrum, and so special care has to be put in designing the filamentary structure. This is indeed the goal of the sGL method: to give an accurate and flexible modeling of the universe as far as its lensing properties are concerned.

\section{Results} 
\label{results}

Our main goal in this Section is to compare the sGL approach with the convergence PDF computed from large scale simulations.
The idea is that by achieving a good agreement with numerical simulations we are proving that the sGL method does provide a good and accurate description of the weak lensing phenomena.
As we shall see, this is indeed the case, as we can naturally reproduce the lensing PDF of the Millenium Simulation.\footnote{We plan to compare the sGL model with other simulations in order to check its accuracy for different cosmological parameters.}

Let us stress that while a comparison to simulations is a good benchmark test for the sGL approach, the simulations and underlying $\Lambda$CDM model do not necessarily describe Nature. Indeed observations do not yet provide strong constraints to the lensing PDF, which leaves room for very different types of large scale structures, examples of which have been studied for example in Ref.~\cite{Kainulainen:2009dw,Amendola:2010ub,Kainulainen:2009sx}.

With the accuracy of the method tested, one can reliably compute the effects of selection biases using sGL. While we lack the necessary expertise to quantitatively estimate the possible bias parameters, we will study a simple toy model for the survival probability function $P^{\rm sur}_{iu}$ to show qualitatively how selection effets might bias the observable PDF.
Finally, we will also show how a JDEM-like survey could constrain the lensing PDF relative to a given cosmological model.
We shall begin, however, by the comparison with the simulations.

\subsection{Comparison with the Millennium Simulation} 
\label{cms}

We shall now confront the sGL method with the cosmology described by the Millennium Simulation (MS)~\cite{Springel:2005nw}. Accordingly, we will fix the cosmological parameters to
$h=0.73$, $\Omega_{M0}=0.25$, $\Omega_{Q0}=0.75$, $w=-1$, $\Omega_{B0}=0.045$ $\sigma_{8}=0.9$ and $n_{s}=1$. Moreover, the mass function of Eq.~(\ref{jenk}) agrees with the MS results. These parameters completely fix the background cosmology and the halos.

We are then left with the filament parameters to specify. First, we fix $\beta_f=0.5$, which means that half of the unvirialized mass forms filaments, while the other half is uniformly distributed. This value determines the maximum possible demagnification $\Delta m_{UE}$ in the lensing PDF, corresponding to  light that misses all the inhomogeneities. For the background model described above this implies $\Delta m_{UE} \simeq 0.17$ mag.

\begin{table}[t]
\caption{\label{param} Filament parameterizations.}
\begin{ruledtabular}
\begin{tabular}{lcc}
Parameters                                                                  &   I   & II                    \\ 
\hline
$\Delta_{f0}$                                                            & 4.5  &  9         \\
$R_{p0}$ ($h^{-1}$Mpc)                                         & 2  & 4                        \\
$L_{p0}$ ($h^{-1}$Mpc)                                           & 20  & 25                        \\
profile                                                                            & uniform  & gaussian                        \\
$M_{f0}^{D}$ ($h^{-1} \Omega_{M0}  M_{\odot}$)        & $9.3 \times 10^{14}$  & $9.3 \times 10^{15}$                        \\      
\end{tabular}
\end{ruledtabular}
\end{table}

The next parameter to set is the present-day filament mass $M_{f0}$. Since $\beta_f$ was taken a constant, a lower $M_{f0}$ implies a higher comoving filament density (see Eq.~(\ref{eq:fila})), a more homogeneous universe and hence less pronounced lensing effects.
In order to connect with the literature we actually have to use the dressed filament mass of Eq.~(\ref{dressed}) which, because of the chosen $\beta_f$, is roughly $M_{f0}^{D}  \approx 3 M_{f0}$.
To choose the parameters defining $M_{f0}$, we follow Ref.~\cite{Colberg:2004cd}, according to which it is reasonable to fix the average filament overdensity to $\Delta_{f0}=4.5$ and the filament radius to $R_{p0}= 2 \, h^{-1}$Mpc.
Note that the overdensity of the dressed filament is roughly $3 \Delta_{f0}$ and that our filaments include also the large halos at the filament ends.
Moreover the density profile seems rather uniform within $R_{p0}$ according to~\cite{Colberg:2004cd} and so we use the simple uniform density profile.
The distribution of filament lengths is quite wide (see Fig.~7 of Ref.~\cite{Colberg:2004cd}) and we choose to use a representative value of $L_{p0}= 20 \, h^{-1}$Mpc. With these choices the bare filament mass becomes $M_{f0}=3.1 \times 10^{14} \, h^{-1} \Omega_{M0}  M_{\odot}$.
We have summarized in Table \ref{param} this first filament parameterization (I).
We can now compare the properties of the lensing PDF from the sGL method to the one from the Millenium Simulation, which is a courtesy of Refs.~\cite{Hilbert:2007ny, Hilbert:2007jd}.

\begin{figure}
\begin{center}
\includegraphics[width=8.5 cm]{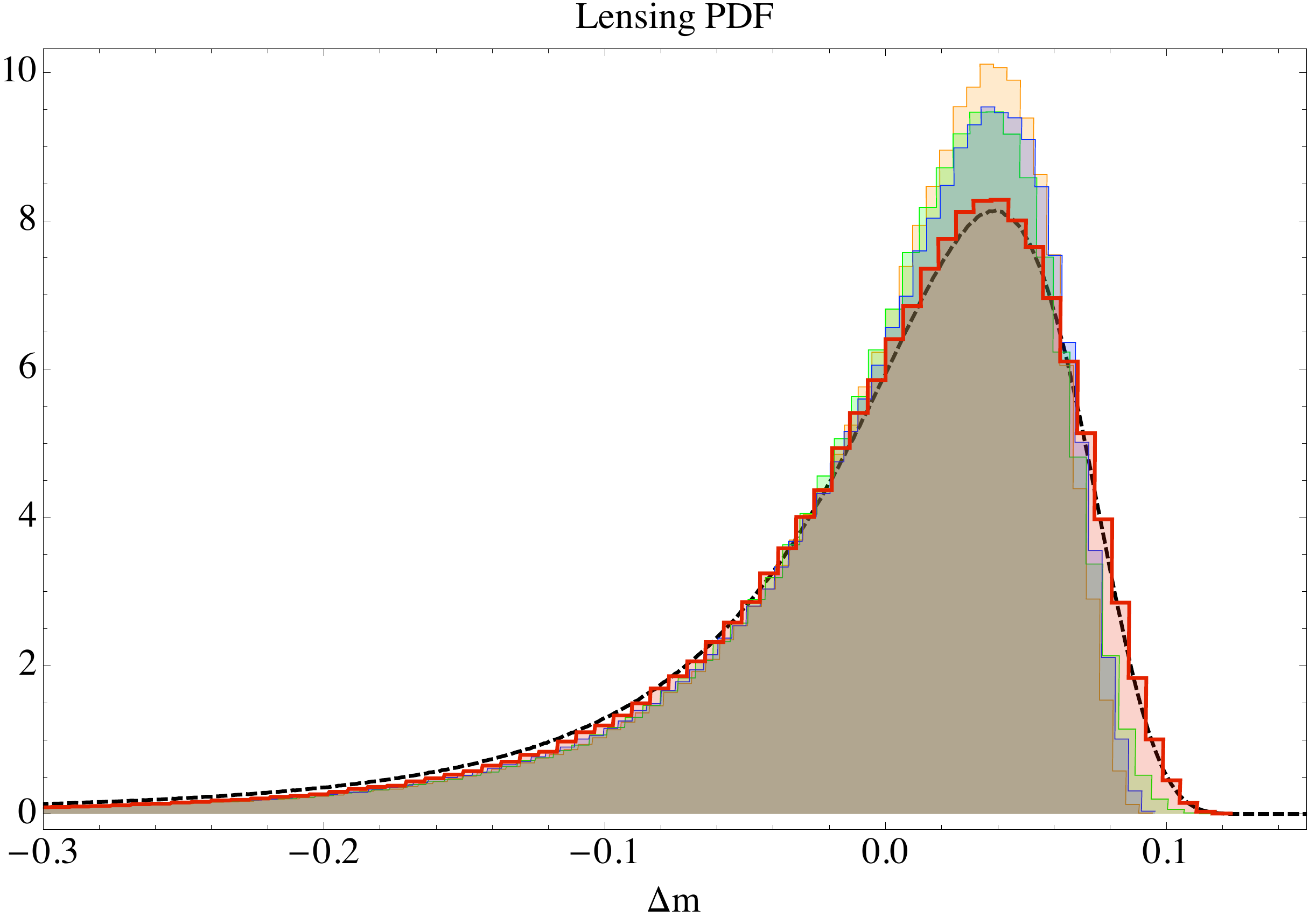}
\caption{Shown are the lensing PDFs for a source at $z=1.5$ for the $\Lambda$CDM model of the Millennium Simulation \cite{Springel:2005nw}.
The dashed line is the lensing PDF generated by shooting rays through the MS \cite{Hilbert:2007ny, Hilbert:2007jd}, while the histograms are obtained with the sGL method. The first (orange) histogram from the top corresponds to a model universe with halos but no filaments, the second (green) is relative to a universe with halos and cylindrical filaments randomly placed and the third (blue) to the case in which the halos are confined within massless filaments. The fourth (red) histogram is the most realistic case with halos confined within randomly placed massive filaments.
The filaments are modelled according to Table \ref{param} (parameterization I). See Section \ref{cms} for details.}
\label{PDFs}
\end{center}
\end{figure}

The four histograms in Fig.~\ref{PDFs} show the four steps in building the model universe in the sGL method. The first (orange) histogram from the top corresponds to a universe endowed only with the halos specified by the mass function.  The second (green) histogram from the top-left represents a model in which half of the uniform matter density ($\beta_f=0.5$) is condensed into filaments, but halos are not confined to them.
The third (blue) histogram from the top-left represents a model in which the halos are confined within the filaments but these are massless.
Finally, the lowest (red) histogram represents the case where the halos are confined within massive filaments.

For these parameters the lensing PDF seems dominated by the halo contribution due to the NFW density profiles and the effect of the large scale clustering seems to increase the dispersion. In particular, the clustering of halos and of smooth mass within filaments seems to give a similar contribution as the two corresponding histograms are close to each other, with a slightly more skewed PDF for the former as compared to the latter. In other words, the nonlinear clustering of halos at the filament scales appears to increase the dispersion similarly to the distribution of smooth matter in low density filamentary structures. The results relative to the Millennium Simulation from Ref.~\cite{Hilbert:2007ny, Hilbert:2007jd} are shown with a dashed line. The agreement becomes very good when the large scale structures are fully included, showing that the lensing PDF is a sensitive probe of the large scale structures. We can conclude that the halos have to be confined within massive filaments if we are to reproduce the MS results, at least for the parameters chosen. Moreover, by fixing the other parameters one could place an upper bound on filament masses. Indeed, by considering the same modeling of parameterization I but with $\beta_f=1$ and $\Delta_{f0}=9$, we have a universe model with the same geometrical properties\footnote{The average filament volume fraction is unaltered if the ratio $\beta_f / \Delta_{f0}$ is kept constant.} but with twice as massive filaments. The resulting lensing PDF has now a longer low-magnification tail ($\Delta m>0$) and is not in good agreement with the MS result. Similarly, we could look for degeneracies between background parameters and filament parameters. For example, still considering the parameterization I, we have found that a lower value of $\sigma_{8}=0.8$ is compensated to a good degree by using a gaussian filament profile instead of a uniform one.

\begin{figure}
\begin{center}
\includegraphics[width=8.5 cm]{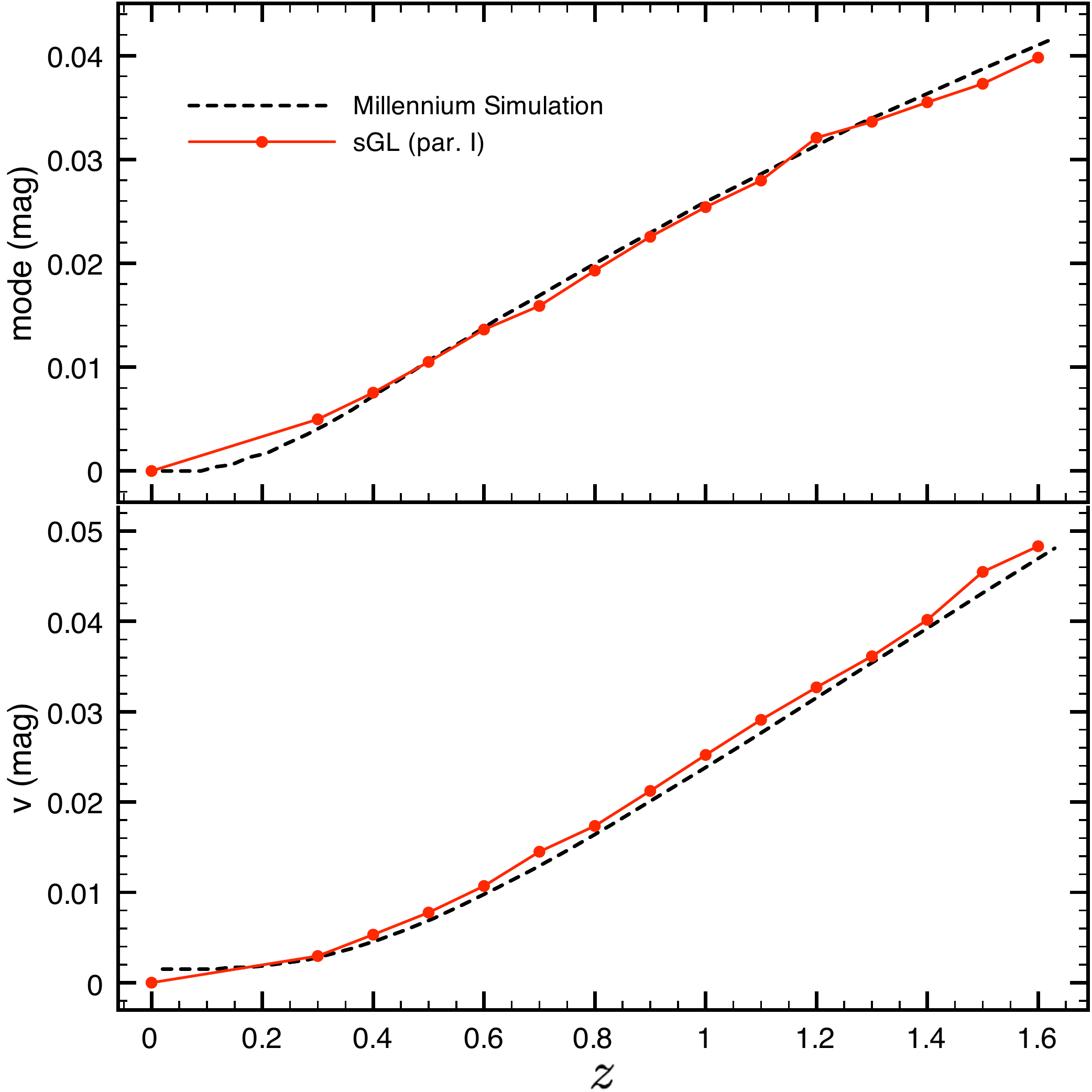}
\caption{Shown is the redshift dependence of mode and dispersion $v=\textrm{FWHM} / 2.35$ of the lensing PDF for the $\Lambda$CDM model of the Millennium Simulation \cite{Springel:2005nw}.
The datapoints were obtained with the sGL method (parameterization I of Table \ref{param}), while the dashed line refers to lensing PDFs generated by shooting rays through the Millennium Simulation (MS) \cite{Hilbert:2007ny, Hilbert:2007jd}.}
\label{modedis}
\end{center}
\end{figure}

The sGL method is an ideal tool to explore the changes induced in the lensing PDF for different choices for the filament parameters and in Fig.~\ref{PDFs2} we show the lensing PDF for a universe in which the filaments are ten times more massive. The same parameters were used as before but now with $\Delta_{f0}=9$, $R_{p0}= 4 \, h^{-1}$Mpc and $L_{p0}= 25 \, h^{-1}$Mpc.
Moreover the filaments have the gaussian profile of Eq.~(\ref{filagau}) and (\ref{gammat}). These parameters are again summarized in Table \ref{param} (parameterization II). It is clear that the agreement with the MS results is lost in this case. The poor agreement, however, does not mean that the universe cannot contain such large structures, but merely that the MS does not have them. 
We will further discuss this example in the next Section.

We have also compared the redshift dependence of the mode and the dispersion of the lensing PDF with the MS results~\cite{Hilbert:2007ny, Hilbert:2007jd} using the filament parameterization I. We choose the full width at half maximum (FWHM) as an indicator for the dispersion instead of the standard deviation $\sigma$ and, to be precise, we will compute the quantity $v=\textrm{FWHM} / 2.35$. The motivation for this is that $v$ is insensitive to the long high-magnification tail and so represents a more robust estimate of the dispersion in the region where the lensing PDF is effectively nonzero (i.e., in the very weak lensing regime). Moreover, for a gaussian PDF one finds $v=\sigma$. The results are shown in Fig.~\ref{modedis}, and the agreement with the results of Ref.~\cite{Hilbert:2007ny, Hilbert:2007jd} is again very good. These results are very encouraging because we next want to study quantitatively the selection effects using the sGL method.

\subsection{Selection Bias Effects}

\begin{figure}
\begin{center}
\includegraphics[width=8.5 cm]{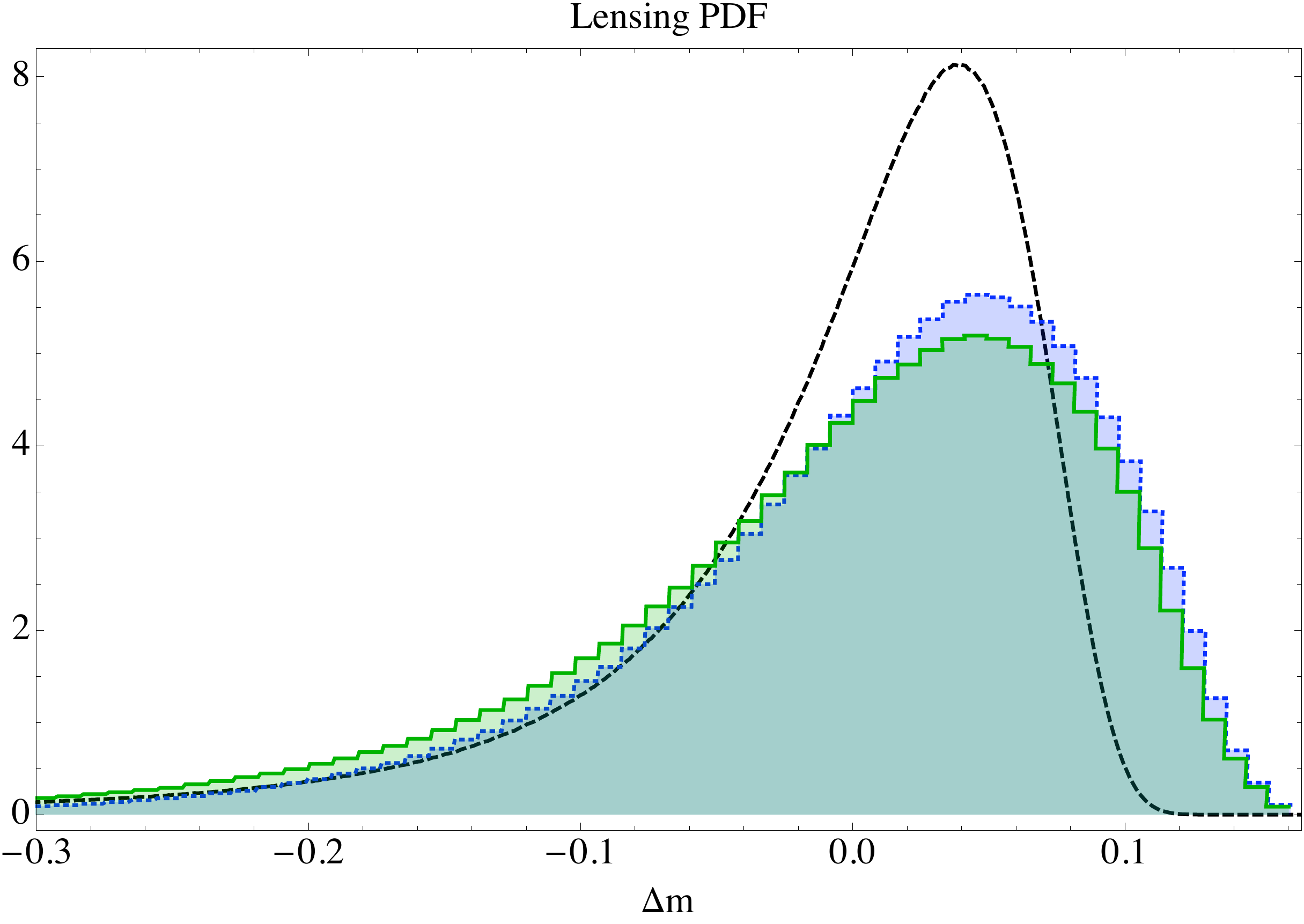}
\caption{Lensing PDF for a source at $z=1.5$ for the filament parameterization II of Table \ref{param}.
The solid (green) histogram is without selection effects, while the dotted (blue) histogram is with the selection effects toy-modelled by Eq.~(\ref{toysel}).
For comparison with Fig.~\ref{PDFs}, the lensing PDF relative to the Millennium Simulation is plotted again as a dashed line.
See Section \ref{results} for details.}
\label{PDFs2}
\end{center}
\end{figure}

\begin{figure}
\begin{center}
\includegraphics[width=8.5 cm]{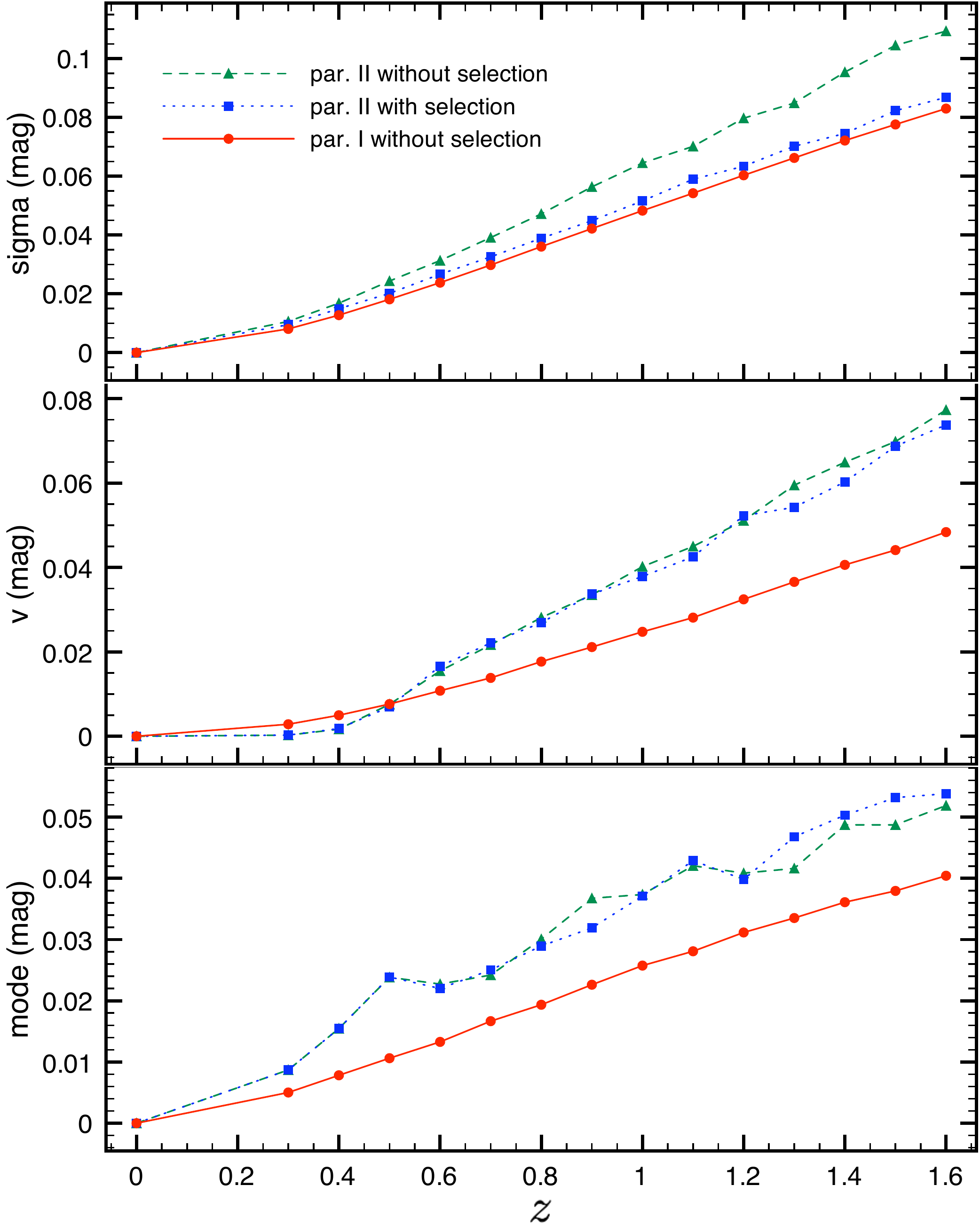}
\caption{Shown is the redshift dependence of $\sigma$ (top panel), $v=\textrm{FWHM} / 2.35$ (middle panel) and mode (bottom panel) for the filament parameterization II of Table \ref{param} with (dotted line) and without (dashed line) the selection effects of Eq.~(\ref{toysel}).
The solid line is relative to the filament parameterization I without selection effects.}
\label{disel}
\end{center}
\end{figure}

We now introduce a toy model for the selection effects to evaluate qualitatively their impact on the lensing PDF.
We model the survival probability simply as a step function in the impact parameter:
\begin{equation} \label{toysel}
P_{\rm sur}(b)= \left\{
  \begin{array}{ll}
    0 & \; b/R< s_{\rm cut} \\
    1 & \; {\rm otherwise} 
  \end{array}\right.\; ,
\end{equation}
where $s_{\rm cut}$ gives the opaque fraction of the halo radius.
In Fig.~\ref{PDFs2} we show the lensing PDF for the filament parameterization II in the case of $s_{\rm cut}=0.1$ (dotted PDF).
The lensing PDF with unitary survival probability is also shown for comparison (solid PDF).
As expected the selection effects reduce the high magnification tail of the PDF.
The effect is more clearly seen by plotting the dispersion as function of redshift for the two cases.
The top panel of Fig.~\ref{disel} shows indeed that the standard deviation $\sigma$ is appreciably smaller when the selection effects are turned on.
The dispersion $v$ (middle panel) and the mode (bottom panel) are, instead, much less affected.

The conclusions are twofold.
First, this shows that $v$ is indeed more robust than $\sigma$ with respect to the high magnification tail, which is effectively cut by the selection effects.
Moreover, the value of $\sigma$ seems to move to the value of $v$ when the survival probability is nontrivial.

Second, while selection effects may reduce the variance, larger scale inhomogeneities have the opposite tendency and the two can compensate each other.
We illustrate this issue by also plotting in Fig.~\ref{disel} the results relative to parameterization I without selection effects.
Clearly the more massive filaments of the parameterization II with selection effects give an effective $\sigma$ very close to the one of the lighter filaments of the parameterization I without selection effects. In other words, for a given observational bound on the lensing variance, there may be some degeneracy between observational biases and weak lensing by large scale structures. This clearly shows the importance of correctly modeling selection effects.
The above degeneracy, however, is broken by the redshift dependence of $v$ and mode, showing that a precise measurement of the lensing PDF (beyond the variance) can yield important cosmological information.
We will discuss in the next Section the possibility of observing the lensing PDF.

\subsection{Measuring the Lensing PDF} \label{obsPDFs}

\begin{figure}
\begin{center}
\includegraphics[width=8.5 cm]{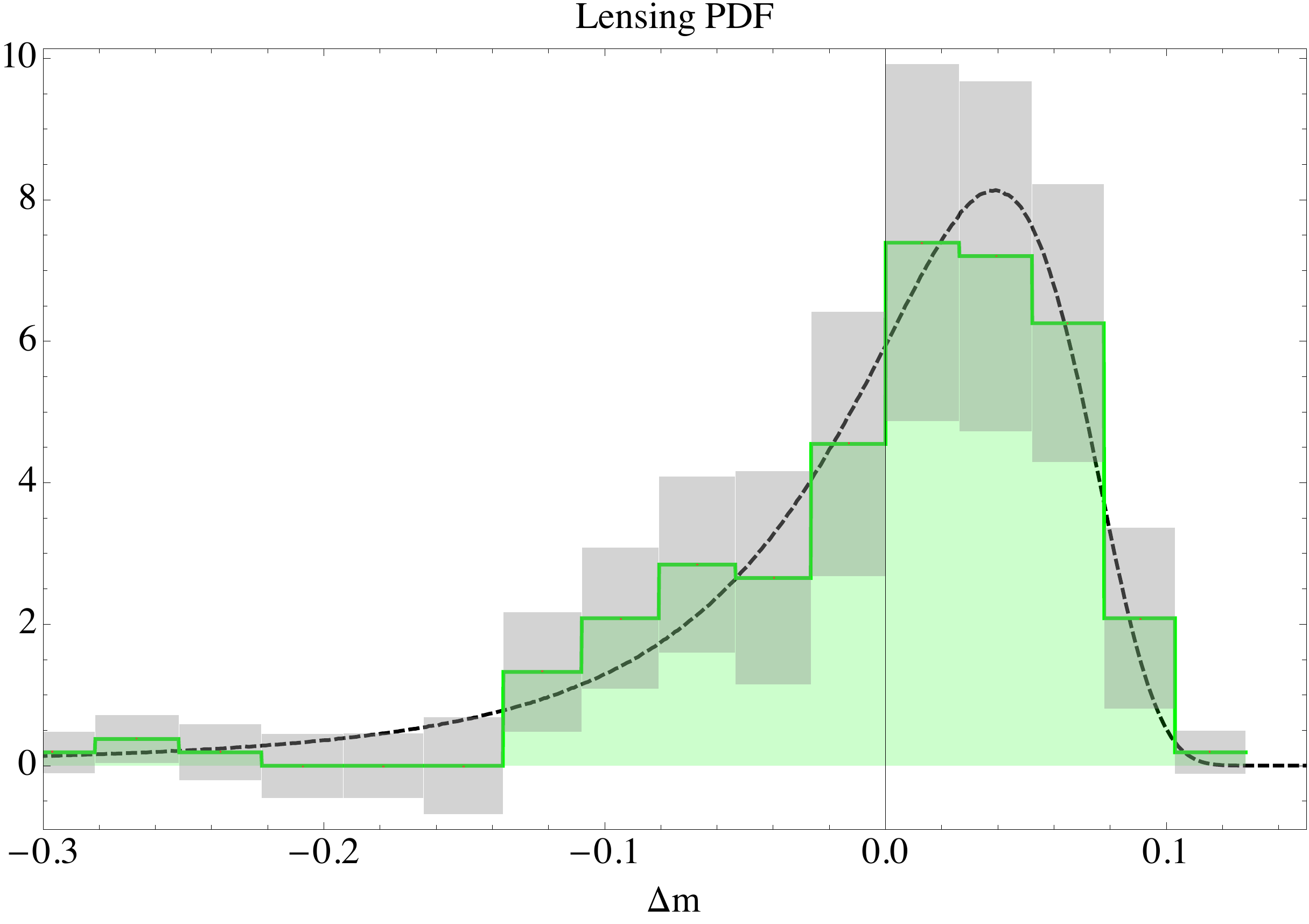}
\caption{Lensing PDF for a source at $z=1.5$ for the filament parameterization I of Table \ref{param} (without selection effects).
The histogram simulates a possible PDF obtained with a dataset of 200 perfect standard candles.
The $2 \sigma$ errors are represented as rectangles.
The PDF relative to the Millennium Simulation is plotted as a dashed line.
See Section \ref{obsPDFs} for details.}
\label{obsPDF1}
\end{center}
\end{figure}

\begin{figure}
\begin{center}
\includegraphics[width=8.5 cm]{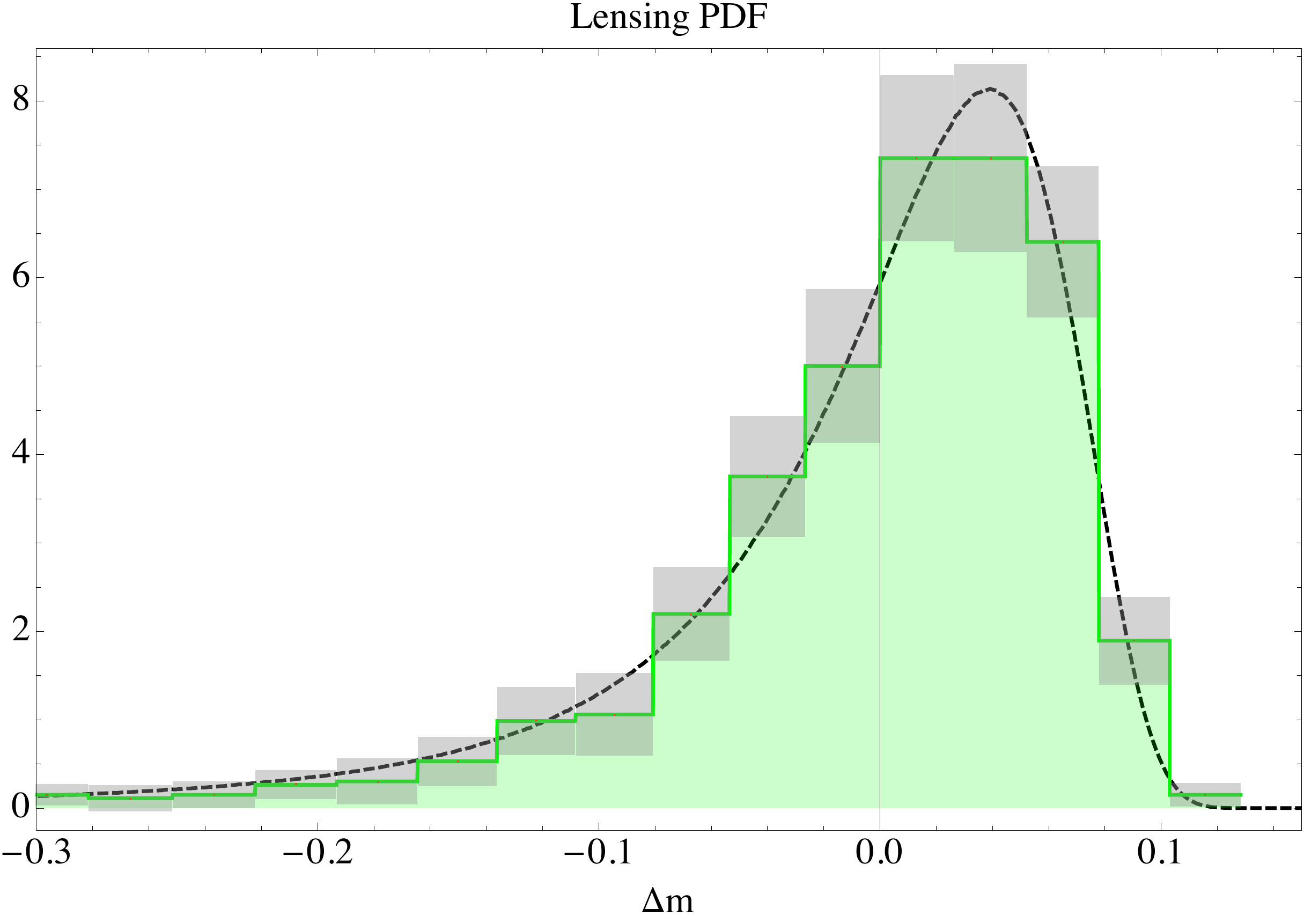}
\caption{Same as in Fig.~\ref{obsPDF1} but for a dataset of 1000 perfect standard candles.}
\label{obsPDF2}
\end{center}
\end{figure}

We will now discuss the possibility of measuring the lensing PDF for the universe of the Millennium Simulation. We will perform a simplified analysis in which we assume that we are observing perfect standard candles and we are thus neglecting the intrinsic dispersion in the SNe absolute luminosity. Moreover, we will also neglect any observational uncertainties. We refer to subsequent work for a more comprehensive analysis, while here we give an illustrative picture for the prospects of observationally constraining the lensing PDF. We will in particular consider a dataset of a JDEM-like survey with one thousand high redshift SNe. With this in mind, we show in Figs.~\ref{obsPDF1}-\ref{obsPDF2} the lensing PDFs for sets of 200 and 1000 perfect standard candles at a given target redshift of $z=1.5$. Also plotted are the $2 \sigma$ errors, where the standard deviation was computed with \mbox{\tt turboGL} in a Monte Carlo fashion by generating many of such histograms and calculating the $\sigma$ for each bin height.

The two figures represent two possible comparisons of the data with a given cosmological model, the difference being the amount of prior cosmological information assumed. Indeed, since the observational data is heterogenous it must be binned to some redshift intervals $\Delta z$ and the SNe luminosities have to be normalized to the same bin redshift.
If $\Delta z$ is small the implicit dependence of the above correction on the background cosmology is small and one finds a {\em direct} constraint on the cosmological model.
This situation is displayed in Fig.~\ref{obsPDF1} for a JDEM-type dataset with 200 SNe in the redshift bin $\Delta z=[1.45, 1.55]$.
It is clear that this single redshift bin cannot put very tight constraints on a generic lensing PDF and one has to use all the high-redshift dataset.

One way to proceed is to compare {\em per bin} the predicted lensing PDF with the observed one, that is, by treating different bins independently.
Another way could be to combine all the 1000 SNe at $z \gtrsim 1$ to one bin.
Fig.~\ref{obsPDF2} displays the relative histogram, which now has much smaller error bars as compared to Fig.~\ref{obsPDF1}.
The normalization of the dataset to the same target redshift, however, used heavily the {\em prior} assumptions about the cosmological model. That is, the comparison shown in Fig.~\ref{obsPDF2} provides only a consistency check for the assumed cosmological model, and it should be noted that the reconstructed lensing PDF might hide a high degree of degeneracy on the cosmological parameters.

\section{Conclusions} 
\label{conco}

In this paper we have revised and extended the stochastic approach to cumulative weak lensing first introduced in Ref.~\cite{Kainulainen:2009dw}. The (sGL) method is based on the weak lensing approximation and on generating stochastic configurations of halos and large-scale structures along the line of sight to cosmological objects, such as supernovae. The halos are intended to model the virialized structures, such as galaxies and clusters of galaxies. We model these structures accurately by means of realistic mass distribution functions and by NFW density profiles (with appropriate $z$ and $M$ dependent concentration parameter). The filamentary structures were modelled with (possibly) non-uniform cylindrical objects.

This modeling incorporates the essential fact that most of the spatial volume is occupied by underdensities (voids), while most of the mass lies in overdense regions of much smaller total volume, in the form of virialized clusters and large filamentary structures. As a consequence the column mass density along a random single geodesic is likely to be lower than the average, which gives rise to the skewness of the lensing PDF. A skewed PDF is of potential importance in particular for the interpretation of the supernova observations, which are still probing much smaller angular scales than the scale at which the homogeneity is recovered. The sGL method with its excellent performance provides a useful tool to investigate the impact of lensing, especially on analyses based on likelihood approaches as done in Ref.~\cite{Amendola:2010ub}.

We confronted our sGL modeling with the lensing results~\cite{Hilbert:2007ny, Hilbert:2007jd} from the Millennium Simulation~\cite{Springel:2005nw}. Even though the MS does not necessarily represent the actual universe, this comparison provides a very useful benchmark test, because here the results are exactly known, as well as the main characteristics of the inhomogeneities involved. The sGL-generated PDF agreed with the MS results very well and naturally without any fine tuning, when all model parameters were chosen to match the typical structures seen in numerical simulations. In the MS cosmology  most of the lensing was found to be due to the halos, with a smaller, but clearly distinguishable contribution coming from filaments. We conclude that it is necessary to embed the halos along massive filaments to reproduce the MS results.

Moreover, the sGL method can easily incorporate observational biases such as selection effects or any other mechanisms that can obscure lines of sight.
We believe this to be an appealing feature of the sGL method, especially with the growing size of SNe datasets: a large number of observations indeed reduces the statistical bias due to the skewed lensing PDF, but does not eliminate the systematic ones possibly caused by selection effects. One can alleviate such problems with a good understanding and modeling of the biases, which is also needed if one wants to observationally constrain the lensing PDF. Indeed, we have shown by a simple toy-model example that selection effects can potentially alter the observable PDF in such a way as to cancel the opposing modification coming from very large scale structures. Assuming no selection effects might then lead to a premature exclusion of such structures.

Finally, we have also shown how a JDEM-like survey could constrain the lensing PDF relative to the cosmology of the Millennium Simulation. Our results suggest that such a survey might be able to give important information about the PDF and thus about the large-scale structure of the universe.

Along with this paper, we release an updated version of the \mbox{\tt turboGL} package, a simple and very fast Mathematica implementation of the sGL method, which can be found at \mbox{turboGL.org}.

\begin{acknowledgments}

It is a pleasure to thank Stefan Hilbert for providing the results of Ref.~\cite{Hilbert:2007ny, Hilbert:2007jd} accompanied by helpful explanations and Gerard Lemson for help in exploring the MS database. The authors benefited from discussions with Luca Amendola, Pierre Astier, Matthias Bartelmann, Stefano Borgani, Julien Guy, Miguel Quartin, Eduardo Rozzo, Barbara Sartoris and Esra Tigrak.

\end{acknowledgments}

\appendix

\section{Performance of the \MakeLowercase{s}GL Method}\label{perf}

In order to estimate the performance of the sGL method we have to calculate the total number $N_{bins}$ of $\{ iu \}$-bins. It is easy to find that
${N_{bins} / N_{S} N_{B}} =   N_{H}   +  N_{F}  N_{T}$,
where $N_{H}$ is the number of halo-mass bins, $N_{F}$ is the number of filaments, $N_{T}$ is the number of $\theta$-bins and we have used for simplicity the same number of impact-parameter bins $N_{B}$ for both halos and filaments.
We remind that $N_{T}$ is the effective number of $\theta$-bins, while the actual number of configurations is $2 N_{T}$.
\begin{table}[h]
\caption{\label{turbo} Performance of the sGL method.}
\begin{ruledtabular}
\begin{tabular}{lccc}
Model type         &   $N_{H}$   & $N_{F}$ & CPU time                       \\ 
\hline
Single-mass halo model & 1  & 0 & 0.2 s        \\
Mass function $f(M)$          & 6  & 0  & 1.1 s                       \\
$f(M)$ + filaments & 6  & 1   & 1.8 s         
\end{tabular}
\end{ruledtabular}
\end{table}
For each $N_{bins}$ the sGL method evaluates the corresponding occupational number from the Poisson statistics.
The program Mathematica using one core of a CPU at $2.5-3$~GHz takes a time $t_{bin}\simeq10^{-3}$~s to produce an array of $N_{stat}=10^{4}$ Poisson numbers.
Other programs will likely have similar performances.
$N_{stat}$ determines the statistics with which the lensing PDF is generated.
Using this information we can then estimate the performance of the sGL method by evaluating $t_{bin} \, N_{bins}$.
The results are shown in Table~\ref{turbo} where we fixed $N_{S}=12$, $N_{B}=15$ and $N_{T}=4$.
Confining halos within filaments typically makes calculations about ten times longer.



\end{document}